\documentclass[AMA,Times1COL]{WileyNJDv5} %STIX1COL,STIX2COL,STIXSMALL

\articletype{Special Issue Paper}%

\received{Date Month Year}
\revised{Date Month Year}
\accepted{Date Month Year}
\usepackage{bm}

\begin{document}
\title{Model bias identification for Bayesian calibration of stochastic digital twins of bridges}

\author[1]{Daniel Andrés Arcones}

\author[2]{Martin Weiser}

\author[3]{Phaedon-Stelios Koutsourelakis}

\author[1]{Jörg F. Unger}

\authormark{ANDRÉS ARCONES \textsc{et al.}}
\titlemark{MODEL BIAS IDENTIFICATION FOR BAYESIAN CALIBRATION OF STOCHASTIC DIGITAL TWINS OF BRIDGES}

\address[1]{\orgdiv{Department 7.7 Modeling and Simulation}, \orgname{Bundesanstalt für Materialforschung und -prüfung}, \orgaddress{\state{Berlin}, \country{Germany}}}

\address[2]{\orgdiv{Modeling and Simulation of Complex Processes}, \orgname{Zuse Institut Berlin}, \orgaddress{\state{Berlin}, \country{Germany}}}

\address[3]{\orgdiv{Professorship for Data-driven Materials Modeling}, \orgname{Technical University of Munich}, \orgaddress{\state{Garching b. München}, \country{Germany}}}

\corres{Daniel Andrés Arcones, Department 7.7 Modeling and Simulation, Bundesanstalt für Materialforschung und -prüfung, Unter den Eichen 87, 12205 Berlin, Germany. \email{daniel.andres-arcones@bam.de}}

\fundingInfo{Priority Program (SPP) 2388/1 "Hundred plus" of the German Research Foundation (Deutsche Forschungsgemeinschaft, DFG) - Project number 501811638.}
%\JELinfo{ejlje}

\abstract[Abstract]{Simulation-based digital twins must provide accurate, robust and reliable digital representations of their physical counterparts. Quantifying the uncertainty in their predictions plays, therefore, a key role in making better-informed decisions that impact the actual system. The update of the simulation model based on data must be then carefully implemented. When applied to complex structures such as bridges, discrepancies between the computational model and the real system appear as model bias, which hinders the trustworthiness of the digital twin and increases its uncertainty. Classical Bayesian updating approaches aiming to infer the model parameters often fail at compensating for such model bias, leading to overconfident and unreliable predictions. In this paper, two alternative model bias identification approaches are evaluated in the context of their applicability to digital twins of bridges. A modularized version of Kennedy and O'Hagan's approach and another one based on Orthogonal Gaussian Processes are compared with the classical Bayesian inference framework in a set of representative benchmarks. Additionally, two novel extensions are proposed for such models: the inclusion of noise-aware kernels and the introduction of additional variables not present in the computational model  through the bias term. The integration of such approaches in the digital twin corrects the predictions, quantifies their uncertainty, estimates noise from unknown physical sources of error and provides further insight into the system by including additional pre-existing information without modifying the computational model.}


\jnlcitation{\cname{%
	\author{Andrés Arcones D},
	\author{Weiser M},
	\author{Koutsourelakis F-S}, and
	\author{Unger, J}}.
\ctitle{Model bias identification for Bayesian calibration of stochastic digital twins of bridges.} \cjournal{\it ASMBI} \cvol{2021;00(00):1--18}.}

\maketitle

\renewcommand\thefootnote{}
\footnotetext{\textbf{Abbreviations:} SHM, structural health monitoring; GP, Gaussian process; OGP, orthogonal Gaussian process; KOH, Kennedy and O'Hagan.}

\renewcommand\thefootnote{\fnsymbol{footnote}}
\setcounter{footnote}{1}

\section{Introduction}

% History/state of the art
The extended availability of computational means in recent years has prompted the digitalization of a plethora of processes across industries. Nowadays, it is not unusual for practitioners, managers and engineers to have large quantities of data at their disposal to enhance their predictions and guide their decisions. Digital twins are key for efficiently harnessing such an amount of information. This is not different in the field of structural engineering and bridge management, where the utilization of digital twins of bridges provides immediate access to the past, current and future state of the structure. Despite recent attempts on its classification\cite{Kukushkin2022}, one of the main challenges for its widespread implementation is the lack of homogenization and standardization in their definition and implementation\cite{Tao2019}. Assuring the fidelity, accuracy and reliability of any particular digital twin and its predictions is therefore crucial in the absence of a common framework within which to evaluate its quality. This holds especially true for applications to structures such as bridges, whose safety and integrity must be preserved under any condition. The objective of this work is to reliably quantify the discrepancy between model and observations in the context of stochastic simulation-based digital twins of bridges to assure their safe and robust deployment for real applications. The main focus will therefore be on those digital twins that use models based on physics and incorporate real sensor data from a structure, e.g. a bridge, representing its current state. They are the so-called "dynamic system models" \cite{Kenett2021,Thelen2022, Thelen2022a}, and their goal is to analyze and monitor the actual state of the structure, for which their predictive power is crucial. Nevertheless, the techniques here implemented are designed to be applicable to any system where a model must replicate a set of observations. Therefore, the additional challenges tied to the dynamic nature of the predictions, in a digital twin \cite{Thelen2022a}, e.g. state-transition uncertainty quantification or real-time computation, remains out of scope.

In contrast to geometry-based digital twins, with a focus on geometrical representation of the information such as Building Information Models (BIM), and purely data-driven digital twins, which focus on sensor observations and monitoring, simulation-based digital twins complement the established knowledge from the system, present in a simulation model, with sensor observations. Following the usual formalization\cite{Carmassi2018, Barbillon2023}, the \textit{simulation} or \textit{computational model} can be represented by a function $f:\mathbb{R}^p\to\mathbb{R}^q$ that depends on a set of parameters $(\bm{\theta},\bm{\lambda})$ that control its behaviour and takes as input a vector of \textit{coordinates} or {variables} $\mathbf{x}\in\mathbb{R}^p$. We differentiate the vector of \textit{latent} parameters $\bm{\theta}\in\mathbb{R}^s$, whose values are unknown beforehand and must be obtained using the observations, from the vector of \textit{prescribed} parameters $\bm{\lambda}\in\mathbb{R}^t$, whose values are set by the modeller beforehand. The real system $z$ is represented by an analogous function that does not depend on $(\bm{\theta},\bm{\lambda})$, as it reflects an infinitely-complex reality without a known parametrization. If we consider a vector of sensor observations $\mathbf{y}_i$ for $\mathbf{x}_i$ from the set of all $n$ collected observations $\hat{\mathbf{y}}$ taken at the set of coordinate vectors $\hat{\mathbf{y}}$, the realization is obtained from the statistical model as
\begin{equation}
	\mathbf{y}_i=z(\mathbf{x}_i)+\epsilon_\text{meas}(\mathbf{x}_i)~~\forall i\in[1,...,n]
\end{equation}
where $\epsilon$ represents a measurement error or \textit{noise} that may depend on $\mathbf{x}_i$. It is generally not possible to access $z$, therefore $f$ is used in its place and the main objective is to obtain the parameters $\bm{\theta}$ best represent the real system. In particular, these predictions often present discrepancies when compared with sensor observations, which reduces their reliability\cite{Chinesta2022}. Considering the computational model $f$, the statistical model results in
\begin{equation}
	\mathbf{y}_i=f(\bm{\theta},\mathbf{x}_i)+\epsilon_\text{model}(\mathbf{x}_i)+\epsilon_\text{meas}(\mathbf{x}_i)~~\forall i\in[1,...,n]
	\label{eq:computational_model}
\end{equation}
where $\epsilon_\text{model}$ refers the so-called \textit{model discrepancy} or \textit{model bias}. The function $y(x)$ will refer to the \textit{generator model} analogous to $z$ that produces the discrete set of observations $\hat{\mathbf{y}}=\lbrace\mathbf{y}_1,\mathbf{y}_2,...,\mathbf{y}_n\rbrace$.

One of the most popular methods to obtain $\theta$ is using Bayesian approaches. Due to its probabilistic nature, uncertainty quantification is also naturally included in the evaluation. The introduction of this uncertainty quantification implies a stochastic component in the predictions and definition of the digital twin. Additionally, as new measurements are obtained, the estimated model can be easily updated to reflect the new reality of the real system. This adaptability is obtained at the expense of often requiring reliable prior information on the potential values of the model parameters and needing numerous evaluations of usually very costly models. For the case of bridges, Structural Health Monitoring (SHM) systems are calibrated following this methodology \cite{Rosafalco2021}, bridge model parameters are obtained using data from measurement campaigns \cite{Rozsas2022} and digital twins are tuned to represent the response of a bridge in real time\cite{Titscher2023}.

Three main challenges arise when implementing a Bayesian framework for digital twins of structures such as bridges in a industrial application: data availability, model complexity and the compatibility between different software pieces and the interfaces that allow the exchange of information between the modules\cite{AndresArcones2023a}. Typically, large quantities of data are required for an accurate model updating, but sensor systems of bridges, despite providing continuous data streams over time, are usually limited to sparsely distributed observation points due to installation and maintenance costs of sensors, as in \cite{Kang2024} for example. Despite attempts to increase the available information through data assimilation approaches \cite{Liu2019a,Liu2019b,Qu2023}, the fidelity of the prediction still depends to a large extent on the data richness and quality. On the other hand, complex models are often unavoidable in the case of simulation-based digital twins. Surrogate models are then introduced as substitutes in time-sensitive applications or to reduce the computational cost for the Bayesian calibration at the expense of introducing further uncertainties\cite{Fu2012,Fu2017, Damblin2018}. The quantification of both surrogate and model discrepancy results critical for such cases. Finally, a large selection of software tools has been developed in recent years that allow for simplified implementations of Bayesian frameworks in larger digital set-ups. Examples of them are \texttt{emcee}\cite{ForemanMackey2013}, \texttt{PyMC}\cite{Salvatier2016}, \texttt{stan}\cite{Carpenter2017}, \texttt{UQpy}\cite{Olivier2020} or \texttt{probeye}\cite{probeye2023}. In this project, the model updating schemes are implemented using \texttt{probeye}, which is an open-source package for Python developed to this end. However, no matter how finely tuned or complex a model is, it will still be limited by its own definition based on a set of prescribed assumptions. The reality, in contrast, is infinitely complex and will never be fully described by a computational model\cite{Oreskes1994}. In the particular case of models updated in a Bayesian framework, the dependency on the training dataset often leads to overfitted and overconfident models that do not reflect the real system \cite{Kaipio2007}. One of the main targets of this work is precisely the quantification of those model discrepancies rooted in deficiencies in the model definition.

For a proper evaluation of the fidelity and reliability of a model, we find it essential to quantify explicitly the discrepancy between the model and the real system. For the purpose of this paper, we tackle the challenges presented by the model discrepancy $\epsilon_\text{model}$ as described in Equation \ref{eq:computational_model} implemented following Kennedy and O'Hagan's framework\cite{Kennedy2001} as an additive term to the model output, leaving the discrepancy at subcomponent level out of scope.

The inclusion of model bias in the inverse problem formulation not only allows quantifying the discrepancy between model and observations, but is also able to improve the inference procedure\cite{Oberpriller2021}. However, the introduction of an overly flexible term to fully capture the model bias incurs potential identifiability issues \cite{Arendt2012}. Throughout the years, several approaches have tried to tackle the identifiability issue, some of which are promising alternatives for their implementation in a digital twin. Brynjardóttir and O'Hagan\cite{Brynjarsdottir2014} highlight the impact of the choice of priors for the bias distribution, outlining the importance of elicitation and expert knowledge in their definition. Modularized versions of Kennedy and O'Hagan's framework\cite{Bayarri2009} attempt to separate the effects of calibrating the model parameter from the bias. Successful examples are its application to energy systems\cite{Menberg2018} and building energy transference estimation\cite{Chong2018}. Non-additive bias structures e.g. using a multiplicative model, are investigated in \cite{Damblin2020, Cocci2022, Cocci2022a} combined with the inclusion of discrepancy terms under Kennedy and O'Hagan's framework in a way that minimizes identifiability problems, which has led to promising results in the field of hydraulics and nuclear engineering. Alternatively, embedded approaches associate additively the bias term with the corresponding latent parameter\cite{Sargsyan2019} or subcomponents of the system\cite{Strong2011}, and have successfully been applied to composite rotor blades\cite{Chen2021}, rocket ignition mechanisms\cite{Sargsyan2019}, jet turbulences \cite{Huan2017} and healthcare policy models\cite{Strong2011, Strong2014}. Finally, modelling the bias term as Orthogonal Gaussian Processes (OGPs) \cite{Plumlee2017} has been proposed as an alternative solution to prevent the identifiability problem between latent and bias parameters while recovering control over the optimality of the process. In this paper, we will compare the suitability of a modularized implementation of the model bias identification framework and the inclusion of OGPs. These two methods are the least intrusive for identifying the model bias, especially in comparison to embedded approaches. This makes them very attractive for digital twins, where modifications of the model are not always possible. To our knowledge, none of these approaches has been implemented or evaluated in the context of digital twins of bridges. Correlated or non-additive bias structures are out of the scope of this paper due to the increased complexity and amount of system information required for their implementation, which may not necessarily be available.

The aim of this study is to implement, assess and extend a model bias identification methodology to be applied to digital twins of bridges, mitigating some of the problems present in Bayesian updating. In particular, we will show that including a bias term corrects the predictions, enhances the model and provides more reliable uncertainty thresholds. Three different approaches will be evaluated: the classical Bayesian inference approach without bias, a modularized version of Kennedy and O'Hagan's framework\cite{Bayarri2009}, and a modification of the same framework to include OGPs in the model bias definition\cite{Plumlee2017}. To further take advantage of the particularities of Digital Twins, two extensions are developed in this work: the quantification of noise from physical sources in the systems response and the extension of the input domain with additional variables. On the one hand, the sensor's network of the Digital Twin often provides a continuous stream of data that is susceptible to present variations due to unaccounted sources. For example, it is not unusual that the deformation sensors installed in bridge monitoring systems have to be compensated due to temperature changes. If the computational model does not take into consideration the temperature, discrepancies between simulations and reality appear \cite{Daro2023}. It is not always possible to identify the source of such discrepancies, which very often appear in the form of noise that must be quantified. The introduction of noise-aware kernels in the bias term enables such a quantification in those cases that lack information of the noise's origin. However, Digital Twins present at their disposal a plethora of documented information sources that can be used to enrich the model. Nevertheless, adapting the computational model to include them typically implies increasing its complexity. Our proposed extension of the model bias term includes the desired additional variables non-intrusively, providing insight into their influence on the model. It is possible, for example, to use the temperature data available for the bridge through its inclusion in the model bias term without needing to modify the computational model. Both extensions will be implemented for the aforementioned approaches of interest in the context of Digital Twins of bridges.

The different approaches will be presented in Section \ref{sec:model_bias_approaches} and applied in Section \ref{sec:applications}. In Section \ref{sub:1dcase}, we will use a simple 1D example to compare how accurate the inferred parameters with each of the approaches are and evaluate their capacity to quantify the model bias. Then, in Section \ref{sub:cantilever_beam}, we will analyze in a stochastic cantilever bridge their ability to quantify the noise from non-prescribed sources and the possibility to include heteroscedastic noise structures with physical meaning. Finally, in Section \ref{sub:bridge}, we will implement each of the approaches to a simplified illustrative model based on a real bridge. We will prove how it is possible to include additional information external to the model, such as a temperature variation, to compensate for the bias in the model. Section \ref{sec:conclusion} closes with the conclusions and potential extensions.

\section{Methodology}
\label{sec:model_bias_approaches}
The methodology takes as base the formulation from Equation \ref{eq:computational_model}. An important modelling choice resides in the structure of the error terms $\epsilon_\text{model}$ and $\epsilon_\text{meas}$. The term $\epsilon_\text{meas}$ defines exclusively the prescribed noise from the sensor. It is generally provided by the manufacturer in the cases of physical sensors or prescribed in the case of virtual ones. Noise will be added to the covariance matrix of the predictions from $z(x, \eta)$. For simplicity, it will be considered uncorrelated Gaussian noise with a prescribed amplitude $\sigma$. The term $\epsilon_\text{model}$ corresponds to the manifestation of the model discrepancy to the observed quantities, which is the object of this study. The reduction of $\epsilon_\text{model}$ is a model identification problem, usually following an iterative approach. In this context, changes to the model are introduced to better match the available information and minimize the discrepancy with the real system. This model update must then consider the available data to identify the fitting parameters and guide potential further improvements in subsequent iterations.

Given a distance $L\left\lbrace y(\cdot)-f(\cdot,\theta)\right\rbrace$ between generating and computational models, the objective of the calibration approaches in a frequentist sense is to find the value of the estimated optimum $\bm{\theta}^*$ that is solution for the problem
\begin{equation}
	\bm{\theta}^*=\underset{\bm{\theta}}{\arg} \min L\left\lbrace y(\cdot)-f(\cdot,\bm{\theta})\right\rbrace.
\end{equation}

This deviation can be quantified in many different ways depending on what is understood as the ``best-fitting" parameters $\theta$. A classical metric for such an approach is the Mean Squared Error (MSE), under which the problem reduces to solving:
\begin{equation}
\min_\theta \frac{1}{n}\sum_{i=1}^n\left(\mathbf{y}_i - f(\mathbf{x}_i, \bm{\theta}) \right)^2.
\end{equation}
This concept can be extended to a functional analysis framework, allowing to define losses over the functions $y$ and $f$ instead of its observations. The main loss considered in this paper is the $L^2$ loss, although other options such as $W_1^2$ can be implemented as well. The $L^2$ loss is generated from the homonymous norm in its associated Hilbert space, being analogous to the MSE in a functional sense\cite{Plumlee2017}:
\begin{equation}
L_{L^2}\left\lbrace y(\cdot)-f(\cdot,\bm{\theta})\right\rbrace := \int_X(y(x)-f(x,\bm{\theta}))^2\text{d}x = \left\| y(\cdot)-f(\cdot,\bm{\theta})\right\| ^2_{L^2}
\label{eq:loss_l2}
\end{equation}
where $X$ is the domain of evaluation of the norm. The evaluation of such a loss in its continuous form is not possible in practice, hence discretization is required. The main difference with MSE is that $L^2$ depends on the discretization of the loss itself while MSE relies exclusively on the observations, which provides more flexibility. This is particularly useful for the case that sparse observations do not correspond with the points of interest, which can make use of alternative discretizations to cover them.

A priori, these metrics would require extensive knowledge about both the generative and the computational model, but they would achieve the best possible fit under a given set of conditions. If the samples represent sufficiently well the domain of the generative process and the predictions are generated with enough information from the computational model, MSE and $L^2$ losses tend to the same value. These frequentist approaches do not inherently provide an uncertainty quantification nor leverage prior information on the parameters. These are desirable qualities in digital twins, therefore Bayesian approaches are favoured.

%%%%%%%%%%%%%%%%%%%%%%%%%%%%%%%%%%%%%%%%%%%%%%%%%%%%%%%%%%%%%%%%%%%%%%%%%%%%%%%%%%%%
\subsection{Bayesian inference without bias}
\label{sub:nobias}

Bayesian inference approaches are based on Bayes theorem\cite{Kaipio2006}
\begin{equation}
p(\bm{\theta}|\hat{\mathbf{y}})=\frac{p(\bm{\theta})p(\hat{\mathbf{y}}|\bm{\theta})}{p(\hat{\mathbf{y}})},
\label{eq:bayes_theorem}
\end{equation}
where $p(\bm{\theta}|\hat{\mathbf{y}})$ is the \textit{posterior} distribution that describes the probability of the parameters $\bm{\theta}$ of having generated $\hat{\mathbf{y}}$, $p(\bm{\theta})$ represent the \textit{prior} distribution based on the beliefs about $\theta$ before seeing the data, $p(\hat{\mathbf{y}}|\theta)$ is the \textit{likelihood} function that represents the probability of observing $\hat{\mathbf{y}}$ for a given $\theta$ and $p(\hat{\mathbf{y}})$ is the \textit{marginal} probability of observing $\hat{\mathbf{y}}$. The prior $p(\bm{\theta})$ must be prescribed, and the likelihood term $p(\hat{\mathbf{y}}|\theta)$ will consider a Gaussian error model on the residuals $\hat{\mathbf{r}}=\hat{\mathbf{y}}-f(\hat{\mathbf{x}},\theta)$ with noise $\epsilon_\text{meas}$. Iterative approaches such as Monte Carlo-Markov Chains (MCMC) converge to an estimation of the posterior distribution up to a constant such that $p(\bm{\theta}|\hat{\mathbf{y}})\propto p(\bm{\theta})p(\hat{\mathbf{y}}|\theta)$, avoiding the calculation of $p(\hat{\mathbf{y}})$\cite{Liu2004}. In this work, an MCMC algorithm with Metropolis-Hastings selection criteria\cite{Kaipio2006} will be applied to the parameter identification problem. This classical approach is well established and admits multiple modifications to improve its efficiency. We differentiate between \textit{unbiased} models, where $\epsilon_\text{model}=0$ is assumed for Equation \ref{eq:computational_model} and \textit{biased} models, where an additive discrepancy term $\epsilon_\text{model}\neq0$ is included. Then, it is necessary to specify a model to quantify $\epsilon_\text{model}$ and its possible dependency on $\mathbf{x}$. 

Within this paper, the parameters $\bm{\lambda}$ will not be explicitly included in the formulation to avoid excessive notation, as they do not play a role in the analysis that will be presented. These are parameters that, despite generally being relevant for the computational model's formulation and evaluation, are prescribed beforehand and are neither inferred nor used as variable input. 

With the generalization of monitoring systems and the ageing of bridges, the necessity to tune a simulation model that represents the bridge behaviour has increased. In most common applications, the available observations correspond to acceleration measurements of a set of sensors dispose over the bridge, coupled with condition sensors such as temperature and humidity\cite{Kang2024}. Then, a dynamic structural model of the bridge is fitted such that the harmonic response obtained from the acceleration corresponds with the model's one \cite{Rozsas2022,Daro2023}. Therefore, the vector of parameters $\bm{\theta}$ corresponds with the geometrical and material properties of the bridge, e.g. Young's modulus $E$, and the target outputs are the eigenfrequencies and associated eigenvectors. Although in some recent cases the model discrepancy has been included in the parameter inference\cite{Conde2018,Barros2023}, the common practice is to increase the complexity of the model until there is a good agreement between predictions and observations\cite{Rozsas2022,Daro2023,Koune2023} either through frequentist or Bayesian inference approaches.

%%%%%%%%%%%%%%%%%%%%%%%%%%%%%%%%%%%%%%%%%%%%%%%%%%%%%%%%%%%%%%%%%%%%%%%%%%%%%%%%
\subsection{Model bias identification approaches}

The first step to correct the shortcomings of classical Bayesian methods regarding the model discrepancy error passes necessarily through acknowledging explicitly its existence. As previously mentioned, it is common practice to include a tunable noise source to absorb and quantify such model deficiencies. In cases, where the model differs greatly from the generative process, the improvements from this approach are limited. Kennedy and O'Hagan introduce in their seminal paper\cite{Kennedy2001} a model discrepancy formulation where the bias term $b$ is explicitly added to the computational model output for a given sample of latent parameters $\bm{\theta}$:
\begin{equation}
\label{eq:bias_koh}
y(x) = f(x,\bm{\theta})+b(x) + \epsilon_\text{meas}(x).
\end{equation}

As the realizations of $f$ are not known a priori, $b$ must be evaluated at some point during the inference procedure. The structure of this term has to be defined beforehand as well, which will influence its behaviour during the parameter tuning. There exist two alternatives on when to evaluate $b$: after inferring the optimal parameter $\theta^*$ or during the inference process. Training $b$ during the parameter tuning further improves the precision and reliability of the resulting parameter $\bm{\theta}^*$\cite{Oberpriller2021}, hence Kennedy and O'Hagan propose fitting $b$ as a Gaussian Process (GP) at every sample of $\theta$ together with the evaluation of the computational model\cite{Kennedy2001}. 

The GP for the bias $b$ is defined by a mean $\mu_b$ and a covariance function $C_b$ which depends on a kernel $k(x,x')$. In most cases, $k(x,x')$ represents a stationary kernel with standard deviation $\sigma_b$ and a length-scale $\ell_b$. Therefore, $b$ is defined as 
\begin{equation}
	b\sim GP(\mu_b,C_b(\sigma_b,\ell_b)).
\end{equation}

We differentiate, consequently, two different responses. On the one hand, we generate the \textit{fitted} response, which returns the predictions $f(x,\bm{\theta})+\epsilon_\text{meas}$ for the inferred posterior distribution of $\bm{\theta}$. This fitted response is a random variable that we will characterize by its mean and variance as
\begin{equation}
	\mu^f=\mathbb{E}\left[f(x,\bm{\theta})+\epsilon_\text{meas}|\hat{\mathbf{y}} \right]
\end{equation}
\begin{equation}
	(\sigma^f)^2=\text{Var}\left(f(x,\bm{\theta})+\epsilon_\text{meas}|\hat{\mathbf{y}}\right).
\end{equation}
This response is bounded by the limitations of the formulation of $f$ and may not be able to represent the observations. On the other hand, we generate the \textit{bias-corrected} response as the output where the bias has been added as,
\begin{equation}
	\mu_\text{dist}^h=\mathbb{E}\left[f(x,\bm{\theta})+b_\theta(x)+\epsilon_\text{meas}|\hat{\mathbf{y}} \right]
\end{equation}
\begin{equation}
	(\sigma_\text{dist}^h)^2=\text{Var}\left(f(x,\bm{\theta})+b_{\bm{\theta}}(x)+\epsilon_\text{meas}|\hat{\mathbf{y}}\right),
\end{equation}
where the bias term has been explicitly indicated as $b_{\bm{\theta}}(x)$ to highlight that each realization depends on the value of $\bm{\theta}$. This correlation of the GP of $b$ and $\bm{\theta}$ presents a caveat for the fully Bayesian setting, as calculating $p(\hat{\mathbf{y}}|\bm{\theta})$ requires marginalizing the GP. Additionally, it is often not practical to evaluate $\mu^h$ and $\sigma^h$ depending on the full distribution of $\bm{\theta}$. Therefore, for practical reasons, the calculation of $b$ will be obtained using an estimation step. Taking inspiration in previous works\cite{Bayarri2009}, we will use a Maximum A Posteriori (MAP) estimation of $\bm{\theta}$ as 
\begin{equation}
	\bm{\theta}^*=\underset{\bm{\theta}}{\arg}\max p(\bm{\theta})p(\hat{\mathbf{y}}|\bm{\theta})
\end{equation}
to calculate the responses that include bias:
\begin{equation}
	\mu^h=\mathbb{E}\left[f(x,\bm{\theta}^*)+b_{\bm{\theta}^*}(x)+\epsilon_\text{meas}|\hat{\mathbf{y}} \right]
\end{equation}
\begin{equation}
	(\sigma^h)^2=\text{Var}\left(f(x,\bm{\theta}^*)+b_{\bm{\theta}^*}(x)+\epsilon_\text{meas}|\hat{\mathbf{y}}\right).
\end{equation}
The resulting final bias GP $b_{\bm{\theta}^*}(x)$ must then be reconstructed from $\bm{\theta}^*$. 

Compared to classical Bayesian inference approaches, the only additional requirement is the need to define and train such GP for $b$. The training process of the GP would depend on the version of the framework to be implemented. This method incurs the same pitfalls of ill-posedness and extreme dependency on the prior's choice. Nevertheless, the additional information provided by the distribution of the bias can be employed to identify deficiencies in the virtual model and its assumptions, as well as to correct the predictions of the computational model based on actual measurements.

Two approaches based on Kennedy and O'Hagan's (KOH) formulation will be implemented in the context of this work: the modularized version of KOH formulated by Bayarri et al.\cite{Bayarri2009} (see Section \ref{sub:mod_koh}) and a formulation based on Orthogonal Gaussian Processes (OGPs) proposed by Plumlee\cite{Plumlee2017} (see Section \ref{sub:ogp}). Both approaches attempt to deal with the identifiability issues that arise with Kennedy and O'Hagan's initial formulation while preserving the predictive capabilities. In both cases, the training process is analogous (see Algorithm \ref{alg:bayarri_kennedy_ohagan_gp}). For a given $\bm{\theta}$, the training input dataset will be $\hat{\bm{x}}$ and the training output will be the corresponding residuals evaluated as the difference between the measurements and the computational model response $\hat{\bm{r}}=\hat{\bm{y}}-f(\bm{\theta},\hat{\bm{x}})$. The parameters of $b$ will be obtained from its maximum a posteriori probability (MAP) estimate as
\begin{equation}
	\mu_b^*,\sigma_b^*,\ell_b^*=\underset{\bm{\theta}}{\arg}\max p(\hat{\bm{r}}|\mu_b,\sigma_b,\ell_b)p(\mu_b,\sigma_b,\ell_b),
\end{equation}
where $p(\hat{\bm{r}}|\mu_b,\sigma_b,\ell_b)$ returns the probability of $b(x,\mu_b,\sigma_b,\ell_b)$ of representing $\hat{\bm{r}}$. This is a well established approach of fitting $b$\cite{Gramacy2020}, although a fully Bayesian approach with Gibbs sampling alternating between samples of $\bm{\theta}$ and $(\mu_b,\sigma_b,\ell_b)$ would be possible but computationally more demanding.

The difference in both methods resides in the definition of the kernels and the covariance matrix of the GP. To our knowledge, they are yet to be applied and evaluated in the context of large structures or simulation-based digital twins. Additionally, these implementations are further extended to identify the noise in biased models and to enhance the model through an extension of the bias' input domain, tackling two of the main shortcomings of classical Bayesian approaches.

%%%%%%%%%%%%%%%%%%%%%%%%%%%%%%%%%%%%%%%%%%%%%%%%%%%%%%%%%%%%%%%%%%%%%%%%%%%%%%%%%%%%%%%%%
\subsubsection{Modularized Kennedy and O'Hagan's formulation}
\label{sub:mod_koh}
In the original KOH's formulation, the GPs for the bias and computational models are trained simultaneously with the inference of $\bm{\theta}$. If any of the components are badly defined due to deficient priors or other error sources, the whole process gets degraded\cite{Bayarri2009}. The contamination between different components can be avoided by modularizing the algorithm. This technique is common practice in engineering systems, and results in a better-defined and less intrusive methodology. A key distinction is the use of the marginal log-likelihood of the fitted GP of $b$ for a given $\bm{\theta}$ as the marginal log-likelihood of the sampled $\bm{\theta}$ itself. As $b$ is defined as a GP, its marginal log-likelihood can be obtained as:
\begin{equation}
	\mathcal{L}(\hat{\mathbf{x}},\bm{\theta})=-\frac{1}{2}\hat{\mathbf{r}}^\top(C_b+\sigma_n^2\mathbf{I})^{-1}\hat{\mathbf{r}}-\frac{1}{2}\log|{C_b+\sigma_n^2\mathbf{I}}|-\frac{n}{2}\log 2\pi
	\label{eq:gp_loglike}
\end{equation}
where the residuals $\hat{\mathbf{r}}$ are defined as $\hat{\mathbf{r}}=\hat{\mathbf{y}}-f(\hat{\mathbf{x}},\bm{\theta})$, and $\sigma_n^2\mathbf{I}$ is the covariance generated by the prescribed measurement noise with amplitude $\sigma_n$ associated with $\epsilon_\text{meas}$. This value is computed using the algorithm 2.1 from Rasmussen\cite{Rasmussen2006}. Therefore, the posterior distribution fulfils
\begin{equation}
	p(\bm{\theta}|\hat{\mathbf{y}})\propto p(b(\hat{\mathbf{x}})|\bm{\theta})p(\bm{\theta})=\exp(\mathcal{L}(\hat{\mathbf{x}},\bm{\theta}))p(\bm{\theta}).
	\label{eq:posterior_modular_koh}
\end{equation}
This approach is the current standard to apply Kennedy and O'Hagan's methodology due to its simple implementation\cite{Gramacy2020}. For those approaches that include a bias term, two different sets of results can be observed. We can generate the posterior predictive distribution for the system as it was done for the case with no bias. As previously mentioned, considering the likelihood of the GP for the bias when evaluating a candidate for $\bm{\theta}$ adds information on the whole generative process, which prevents the posterior predictive from converging to the MSE with respect to the observations. This awareness of the underlying generative process makes the prediction less sensitive to model discrepancy. 

\begin{algorithm}
	\caption{Modularized version of Kennedy and O'Hagan by Bayarri (2009)}
	\label{alg:bayarri_kennedy_ohagan_gp}
	\begin{algorithmic}[1]
		\Procedure{MCMC run}{}
		\For{Evaluation of $\bm{\theta}$ in the MCMC run sampled from the last state of the chain}
		\State Define the bias model $b(x)$ as a GP on the residuals $\hat{\mathbf{r}}=\hat{\mathbf{y}}-f(\hat{\mathbf{x}},\bm{\theta})$ for a given $\bm{\theta}$ 
		\State Fit $b(\hat{\mathbf{x}})$ with $\hat{\mathbf{r}}$
		\State Compute the log-marginal likelihood of the fitted $b$ as in Equation \ref{eq:gp_loglike}
		\State Use the marginal likelihood of $b$ and the prior of $\bm{\theta}$ to obtain the posterior distribution for the sampled $\bm{\theta}$ as in Equation \ref{eq:posterior_modular_koh}
		\State Decide if keeping or rejecting $\bm{\theta}$ based on its prior and likelihood evaluations
		\EndFor
		\EndProcedure
		\State Regenerate the full bias-corrected response $f(x, \bm{\theta}^*)+ b(x)$ for the estimated optimum value $\bm{\theta}^*$.
	\end{algorithmic}
\end{algorithm}
%%%%%%%%%%%%%%%%%%%%%%%%%%%%%%%%%%%%%%%%%%%%%%%%%%%%%%%%%%%%%%%%%%%%%%%%%%%%%%%%%%%%%%%%%
\subsubsection{Orthogonal Gaussian Processes}
\label{sub:ogp}
To mitigate the identifiability problem of approaches based on KOH's formulation, Plumlee proposes modelling the discrepancy term as an Orthogonal Gaussian Process (OGP) \cite{Plumlee2017}. This is achieved by imposing an orthogonality restriction over the bias such that it represents exclusively the discrepancy between observations and predictions that could not be explained by $f$, supposing that the given sampled values for $\bm{\theta}$ were the best-fitting ones. Therefore, the bias is defined as orthogonal to the gradient of a specific loss function, $L^2$ loss in this case, with respect to the latent parameters.

This is implemented as a modification on the prior of the GP for the bias. In the case where $f$ is the original computational model evaluated at the $N$ anchor points $\bm{\xi}_i$, the prior for the bias is defined as $b(x)\sim GP(0,C_b)$, with
\begin{equation}
\label{eq:Cb_ogp}
C_{b_\theta}(x,x')=k_b(x,x')-\bm{h}_\theta(x)^TH_\theta^{-1}\bm{h}_\theta(x')
\end{equation}
where 
\begin{equation}
\bm{h}_\theta(x)=\dfrac{1}{N}\sum_{i=1}^{N}\dfrac{\partial}{\partial\bm{\theta}}f(\bm{\xi}_i,\bm{\theta})k_b(x,\bm{\xi}_i)
\end{equation}
and
\begin{equation}
H_\theta = \dfrac{1}{N^2}\sum_{i=1}^{N}\sum_{j=1}^{N}\dfrac{\partial}{\partial\bm{\theta}}f(\bm{\xi}_i,\theta)\left[ \dfrac{\partial}{\partial\bm{\theta}}f(\bm{\xi}_i,\bm{\theta})\right] ^T k_b(\bm{\xi}_j,\bm{\xi}_i),
\end{equation}
where $k_b$ is the original correlation function prescribed by a kernel and a standard deviation. A subindex $\theta$ has been added, following the original implementation from Plumlee\cite{Plumlee2017}, to emphasise that $C_{b_\theta}$, $\bm{h}_\theta(x)$ and $H_\theta$ are only valid for the sampled $\bm{\theta}$ at which they are evaluated. The anchor points can be defined, for example, as the nodes of a mesh that discretizes the geometry of the system, the training points of a GP or just a set of points where $f$ must be evaluated as a discrete approximation of $f$ to determine the orthogonality condition. In summary, the anchor points correspond to those input values that define the system's response in the domain of interest. In contrast with the other implemented approach, these anchor points do not necessarily coincide with the observation points. For example, given a simulation of a physical system whose geometry is modelled with a discrete mesh, the observation points will be those for which there are measurements, while the anchor points will be all the nodes of the mesh. Therefore, the definition of $\hat{\bm{\xi}}$ introduces a potential difference between the MSE of the observations and the inferred optimum for the $L^2$-norm, which enforces the orthogonality condition over $\hat{\bm{\xi}}$ and not only over the observations. These equations can be expressed in matrix form, being $t$ the dimensionality of $\bm{\theta}$, as 
\begin{equation}
\label{eq:Cb_ogp_matrix}
C_{b_\theta}(x,x')=k_b(x,x')-\bm{w}(x)^TF_\theta\left(F_\theta^TWF_\theta \right)^{-1}F_\theta^T\bm{w}(x')
\end{equation}
where $\bm{w}(x)$ is the $N\times 1$ vector with elements $k_b(x,\bm{\xi}_i)$, $F_\theta$ is the $N\times t$ matrix with rows  $\dfrac{\partial}{\partial\bm{\theta}}f(\bm{\xi}_i,\bm{\theta})$ and $W$ is an $N\times N$ matrix with elements $ k_b(\bm{\xi}_j,\bm{\xi}_i)$.

The introduction of the orthogonality condition mitigates the identifiability issue without requiring a well-defined prior distribution and reduces the variability of the inferred latent parameters\cite{Plumlee2017}. These are valuable advantages in the case of digital twins, where the increased computational costs is only relevant for the off-line phase of the model, where the focus lies on the calibration and update of the model and not on its fast evaluation through new data points.

The inferred parameters with OGP bias tend to converge to the optimal value for the predefined loss and present narrower distributions than the modularized formulation of KOH due to the orthogonality condition. This effect is especially relevant if the optimum value of the latent parameters over the whole system domain is expected to differ largely from the optimum over only the sampled points. However, in contrast to KOH's implementation and its modularized version, OGPs require the evaluation of the derivatives of $f$ with respect to the latent parameters $\bm{\theta}$, which increases its evaluation times.

\subsection{Model bias extensions}
\subsubsection{Identification of noise in biased models}
One relevant application of Bayesian inference schemes is the estimation of latent parameters associated with noise. These noise terms can present \textit{homoscedastic} and \textit{heteroscedastic} structures. Homoscedastic noise structures assign the same amplitude to every observation. In the absence of correlation, this implies a diagonal matrix with identical entries for the covariance. Heteroscedastic noise structures admit different amplitudes for each observation, which are then extended to the full domain through kernel regression. The resulting matrix will still be diagonal, but its entries will vary depending on the observation coordinates.

In general, defining the bias term as a GP process requires the selection of suitable kernels. Typical choices are Radial Basis Functions (RBF) or Matérn kernels\cite{Gramacy2020}, aimed at regression problems. In the implementation of the GP, a prescribed white noise of variance $\sigma_n^2$ is usually added as in Equation \ref{eq:gp_loglike}. It provides numerical stability and, in some cases, represents the measurement error $\epsilon_\text{meas}$. Due to its flexible structure as a GP, the bias term will absorb $\epsilon_\text{meas}$, thus including it explicitly is not required. It is generally not possible to distinguish between $\epsilon_\text{model}$ and the non-prescribed $\epsilon_\text{meas}$. We will then consider any variation in the model discrepancy as part of $\epsilon_\text{model}$ and susceptible to be explained by the bias term. If more than one distinct output observation is available per input, the fitted GP tends to solve the regression problem and will not capture that variability. We consider these variations in the observations as noise to be identified, hence a noise-aware kernel with a non-vanishing variance at the observation points is required in those cases. This concept is analogous to the \textit{nugget effect} considered in geostatistics\cite{Camana2019}. In that context, we formalize a noise-aware kernel $k(x_i,x_j)=k(h)$ where $h$ is the \textit{lag} or distance between $x_i$ and $x_j$, as such for which the semivariogram $\gamma(h)$ \cite{Chiles1999} of the resulting GP returns a non-zero value when $h\to0$:
\begin{equation}
	\lim_{h\to0}\gamma(h)\neq 0.
\end{equation}

Homoscedastic and heteroscedastic noise kernels are available to be added to the GP. Homoscedastic ones, such as white noise kernels, allow us to identify constant noise added to the observations. As the bias term aims to compensate for the discrepancy in the output of the model and $\epsilon_\text{meas}$ is assumed to be prescribed, homoscedastic kernels in the bias GP would capture additional observation noise of non-physical nature, such as deficient data collection. A typical homoscedastic kernel is a white noise kernel with noise variance $\sigma_n^2$ such that
\begin{equation}
	k_{WN}(x_i,x_j)=\begin{cases}
		\sigma_n^2 & \text{if }x_i = x_j\\
		0 & \text{otherwise}
	\end{cases}.
\label{eq:homo_wn}
\end{equation} 
This kernel can be seamlessly added to the regression kernel of the GP bias to provide it with noise-awareness. In practice, it appears as an additional $\sigma_n^2$ in the diagonal of the covariance matrix for the training data points. Therefore, for the bias with an additional white noise kernel,
\begin{equation}
	C_{b+WN}=C_b+\sigma_n^2\mathbf{I}.
\end{equation}

Alternatively, heteroscedastic noise kernels assign different noise values to each observation. Adding such a kernel to the GP for the bias allows quantifying variable noises, which typically come from a physical source, such as a variable material parameter or changing measurement/environmental conditions. An example of a heteroscedastic noise kernel -and the one used throughout this work- is such that for each element in a set of anchor points $\hat{\bm{\xi}}$, an associated noise variance $\sigma_i^2$ is defined. For any other point, the noise variance is obtained through regression on the prescribed values. Therefore, the heteroscedastic noise kernel is defined as
\begin{equation}
	k_{HN}(x_i,x_j)=\begin{cases}
		\sigma_1^2 & \text{if }x_i = x_j = \bm{\xi}_1\\
		\sigma_2^2 & \text{if }x_i = x_j = \bm{\xi}_2\\
		\sigma_3^2 & \text{if }x_i = x_j = \bm{\xi}_3\\
		\dots \\
		g(x_i) & \text{if }x_i=x_j\text{ and }x_i\notin\hat{\bm{\xi}} \\
		0 & \text{otherwise}
	\end{cases}.
\label{eq:hetero_hn}
\end{equation}
where $g(x_i)$ represents the evaluation of the regressor built on the pairs $(\bm{\xi}_i, \sigma_i^2)$. Determining the set of anchor points $\hat{\bm{\xi}}$ is not straightforward in real applications. They control the behaviour of $g(x_i)$: too many anchor points become expensive to compute and may lead to overfitting, while too few may not capture the variation in $\sigma_i^2$. It is general practice to take the training points of the GP as anchor points \cite{Sung2019}. If no additional information is available, this choice reflects the expectations that the training points will capture sufficiently well both bias and noise variance. Further considerations on how to model heteroscedasticity in GPs can be found in the work of Binois et al\cite{Binois2016}.

In the implementation used in this work\cite{Metzen2016}, $g(x_i)$ is obtained from applying kernel regression with the training points as anchor points $\hat{\bm{\xi}}$. This kernel can be added analogously to the homoscedastic one to the regressor kernel of the bias GP, where its contribution to the covariance matrix $C_{HN}$ is presented as a diagonal matrix with different entries obtained from the kernel evaluation:
\begin{equation}
	C_{b+HN}=C_b+C_{HN}
\end{equation}

The introduction of noise-aware kernels in the bias allows quantifying the variability in the observations without modifying the deterministic computational model. These kernels can be implemented in both KOH and OGP approaches without needing to further modify the methodology presented in the previous sections. In the case of OGPs, the orthogonality condition is not affected by the noise introduction, as it is part of the kernel evaluation. Sung et al. \cite{Sung2019} verified that the introduction of heteroscedastic noise with OGP bias improved the calibration of models in comparison with taking homoscedastic noise with no bias or KOH's approach with no noise. In this paper, the comparison is extended to the framework of their applicability to digital twins of real structures.

\subsubsection{Model enhancement through bias' input domain extension}
One of the most prominent advantages of implementing the bias term independently from the model is the possibility of including information in the bias that enriches and guides the model inference without needing to modify the computational model itself. Most approaches that extend a given computational model when bias is present use data-driven models that can be easily modified \cite{Carmassi2018, Salter2019}, iteratively update the models \cite{Damblin2018, Damblin2020, Cocci2022, Cocci2022a}, or modify the model by defining functional outputs\cite{Ma2019}. A review on such approaches can be found in Baker et al. \cite{Baker2022} and Sung et al. \cite{Sung2024} In contrast, the aspect presented here has the potential to enhance the predictions generated by simulation-based digital twins without modifying the model itself, only through a modification of the bias representation.
 
Being $f(x,\theta)$ the computational model, we can extend the model with a bias $b(x,\eta)$ such that $\eta$ are parameter coordinates not considered in $f$ but for which observations are available. Therefore, obviating the measurement noise $\epsilon_\text{meas}$, the modified Equation \ref{eq:computational_model} for the final model would be
\begin{equation}
	z(x,\eta,\theta)=f(x,\theta)+b(x,\eta),
\end{equation}
which would be defined over the extended domain $(x,\eta)\in X\times H$. For example, a temperature-agnostic model defined over the spatial domain $X$ could be extended to consider the temperature by including it in the bias term. This approach requires the datasets used for calibration to be labelled for the extension variable as well as for the original ones. Such is the case for many digital twins, especially for large structures such as bridges, where abundant information from a variety of sources is gathered independently for monitoring purposes but not necessarily included in the model definition.

The introduction of an extension of the domain requires minor modifications to the inference workflow. In particular, the input training data is composed of pairs of $(x,\eta)$, while the computational model is only defined for $x$. Nevertheless, the GP for the bias takes as output training data the residuals, which are defined over $(x,\eta)$, which allows for keeping the associated pairs of $x$ and $\eta$. Therefore, for the training data of the GP, we define as input dataset the pairs $(\hat{\mathbf{x}},\hat{\bm{\eta}})$ and as output dataset the residuals for a given sampled $\bm{\theta}$
\begin{equation}
	\hat{\mathbf{r}}(\hat{\mathbf{x}},\hat{\bm{\eta}},\bm{\theta})=y(\hat{\mathbf{x}},\hat{\bm{\eta}})-f(\hat{\mathbf{x}},\bm{\theta}).
\end{equation}
This extension only requires a modification in the evaluation of the residuals for a given sample of the latent parameters, which propitiates its integration in the developed workflows. Notably, the computational model $f$ would still be defined only over $x$, hence its predictions will not depend on the value of $\eta$. However, the bias-corrected response will vary depending on the additional information included through $\eta$. 

Remarkably, the optimization for the best-fitting bias GP includes the variability introduced in $\eta$. The bias correction then provides information on the deficiencies in the model with respect to $\eta$ and influences the proposed distribution of the optimal $\theta$. This allows to gain insight into the distribution of the model discrepancies in the whole domain $(x,\eta)$ and their associated uncertainties. The results provided by the extension can be used to assess the quality of the model and guide potential future improvements.

It must be noted that the orthogonality condition from OGP is solely dependent on $f$, therefore it includes no information about the extended domain. Consequently, the extension of the domain does not affect the definition of the optimal $\theta$ in that case. The resulting extension of the bias still provides information about the distribution of the discrepancy in $\eta$.

%%%%%%%%%%%%%%%%%%%%%%%%%%%%%%%%%%%%%%%%%%%%%%%%%%%%%%%%%%%%%%%%%%%%%%%%%%%%%%%%%%%%%%%%%%%%%%%%%%%%
\section{Applications}
\label{sec:applications}
\subsection{Bias identification with non-isotropic data densities}
\label{sub:1dcase}
To illustrate the distinct methodologies, a modification of Plumlee's ``Pedagogical example"\cite{Plumlee2017} is implemented. In this case, the impact of non-isotropic data densities, i.e. gaps in the measurements, on the bias identification is evaluated. Table \ref{tab:1d_case} presents a simple 1D case that is suitable for the different approaches. Taking $X=[0, 1]$ as the complete domain of interest, the original response is generated from a "true" model governed by $y(x)= 4x +x\sin 5x$ for $x\in X$, while the computer model is $f(x, \theta)=\theta x$. Measurements are collected for $\hat{X}=[0,0.25]\cup[0.4,0.5]\cup[0.8,1.0]$ at equidistant intervals of 0.05 and are perturbed with Gaussian noise of variance $\sigma^2=0.02$. It is not uncommon that in practice measurements are not available for some parts of the domain of interest, represented in this case by the arbitrarily chosen intervals where no measurements are collected. This ``gaps'' in the measurement collection generates the aforementioned non-isotropic data densities, where some intervals of $X$ are represented with fewer measurements in the training set.

\begin{table}[htb]
	\centering
	\begin{tabular}{rl}
		\hline
		Element & Formulation \\ \hline
		Generator model  & $y(x)= 4x +x\sin 5x$ \\ 
		Computational model without bias term & $f(x, \theta)=\theta x+\epsilon$\\
		Computational model with bias term & $z(x, \theta)=f(x, \theta)+b(x)=\theta x+b(x)+\epsilon$ \\
		Discrepancy source & $x\sin 5x$\\
		Latent parameter and prior & $\theta,~p(\theta)\sim\mathcal{N}(2.5,1.5)$ \\
		Domain of interest  & $x\in X=[0,1]$ \\ 
		Measurements' domain & $\hat{x}\in\hat{X}=[0,0.25]\cup[0.4,0.5]\cup[0.8,1.0]$\\
		Noise & $\epsilon\sim\mathcal{N}(0, \sqrt{0.02})$\\
		\hline
	\end{tabular}
	\caption{Case description for Section \ref{sub:1dcase} (``Pedagogical example" model).}
	\label{tab:1d_case}
\end{table}

\begin{figure}[ht]
	\centerline{\includegraphics[width=0.8\textwidth]{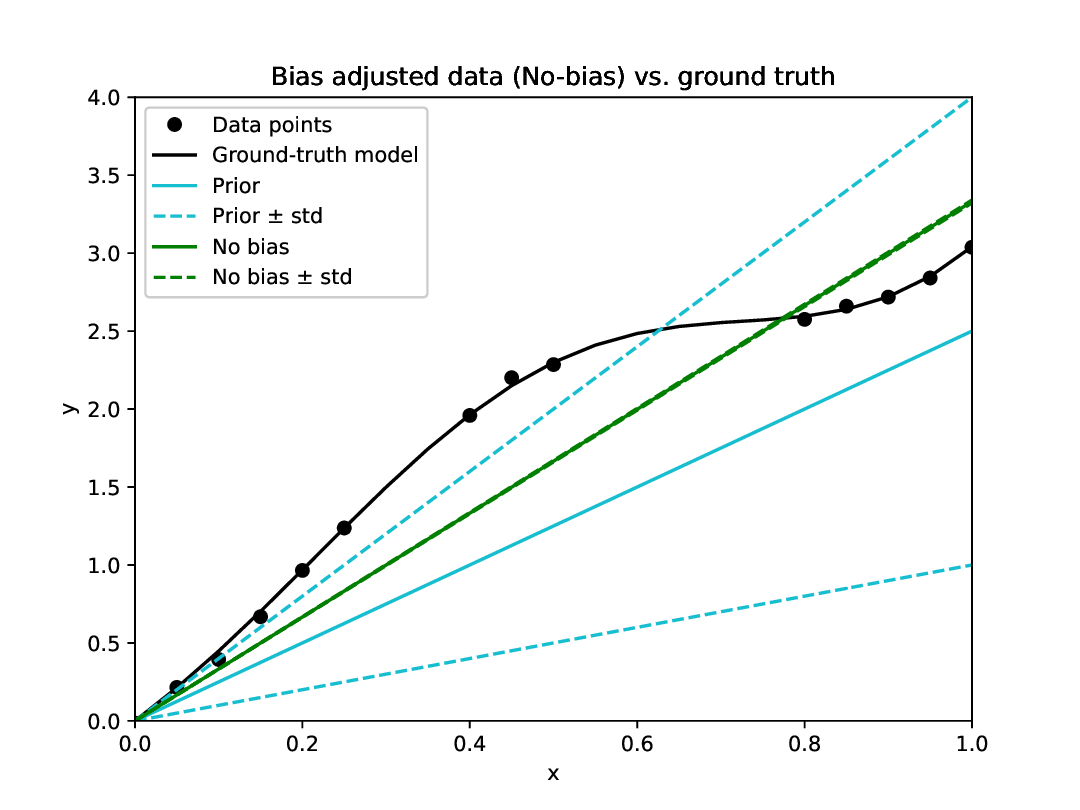}}
	\caption{Fitted system's response for bias-free Bayesian inference, 1D case}
	\label{fig:1d_nobias}
\end{figure}

The objective is to fit the variable $\theta$ such that the computer model reflects the measurements. There is a clear discrepancy between $y$ and $f$, therefore it can potentially be quantified. Our modifications to the original example focus on highlighting the effect of an incomplete dataset and the comparison of the different approaches for the same system. To be used as a baseline model, the classical Bayesian approach without bias is implemented. Therefore, for a given $\theta$ and given evaluations $\hat{\bm{\epsilon}}$ of the Gaussian noise $\epsilon\sim\mathcal{N}(0.0,\sigma)$ for each observation point, the residuals of the predictions at the measurements can be expressed as
\begin{equation}
	\hat{\mathbf{r}}=(\hat{\mathbf{y}}-\theta\hat{\mathbf{x}})+ \hat{\bm{\epsilon}}.
\end{equation}
The posterior distribution for $\theta$ is estimated using Bayesian inference. The prior for $\theta$ is modelled as a normal distribution $\theta\sim \mathcal{N}(2.5,1.5)$. This prior represents a poor initial belief with a high degree of uncertainty that still includes the observations in a reasonable range (see Figure \ref{fig:1d_nobias}). A Gaussian model with prescribed additive error of $\sigma^2$ is defined as proposal distribution. The MCMC algorithm is run for 1000 steps with a burn-in of 100 steps, which is enough to ensure the chain convergence as shown in Table \ref{tab:Simple_inference}. The posterior predictive resulting from sampling the inferred $\theta$ is represented in Figure \ref{fig:1d_nobias}. 
\begin{center}
\begin{table*}[!h]
	\centering
\begin{tabular}{lccccccc}
\toprule
                 Group &     Mean &        SD &    HDI$_{3\%}$ &   HDI$_{97\%}$ &  MCSE$_{\text{mean}}$ &   MCSE$_{\text{SD}}$ &     $\hat{R}$ \\
\midrule
              No bias &   3.33 & 0.01 &    3.31 &     3.34 &       0.00 &     0.00 &    1.01 \\
KOH &  3.37 & 0.34 &    2.70 &     3.98 &       0.01 &     0.01 &    1.03 \\
 
OGP  &   3.52 & 0.04 &    3.44 &     3.60 &       0.00 &     0.00 &    1.02 \\
\bottomrule
\end{tabular}
\caption{Summary inference results for parameter $\theta$ in the simple case benchmark. Mean and SD are the statistics of the sampled parameters. HDI 3\% and 97\% are the bounds of the 94\% Highest Density Interval. MCSE$_\text{mean}$ and MCSE$_\text{SD}$ are the Monte Carlo Standard Errors due to having a finite number of samples in the MCMC for the mean and standard deviation, respectively. $\hat{R}$ is the Gelman-Rubin diagnostic\cite{Gelman1992} (1 = perfect convergence).}
\label{tab:Simple_inference}
\end{table*}
\end{center}

It can be observed that the prediction has converged close to a Dirac's delta distribution due to the measurement noise being prescribed with a small variance. The model cannot represent the measurements because of model discrepancy. Uncertainty is also underestimated due to such discrepancy. Additionally, the unbiased approach provides a value of $\theta$ that makes the predictions at the observation points converge to the minimum Mean Square Error (MSE) with respect to the measurements, with a value of 3.34 and a standard deviation of 0.01. For reference, the optimal value of $\theta$ obtained using an $L^2$ loss (see Equation \ref{eq:loss_l2}) is 3.6. The value obtained does not coincide with the $L^2$ norm between the generating process and the computational model over the domain of interest $X=[0,1]$ due to undersampling of the regions $[0.25,0.4]$ and $[0.5,0.8]$.  This discrepancy further reduces the reliability of the inferred computational model far from the observation points, which is undesirable.

\begin{figure}[ht]
	\centerline{\includegraphics[width=\textwidth]{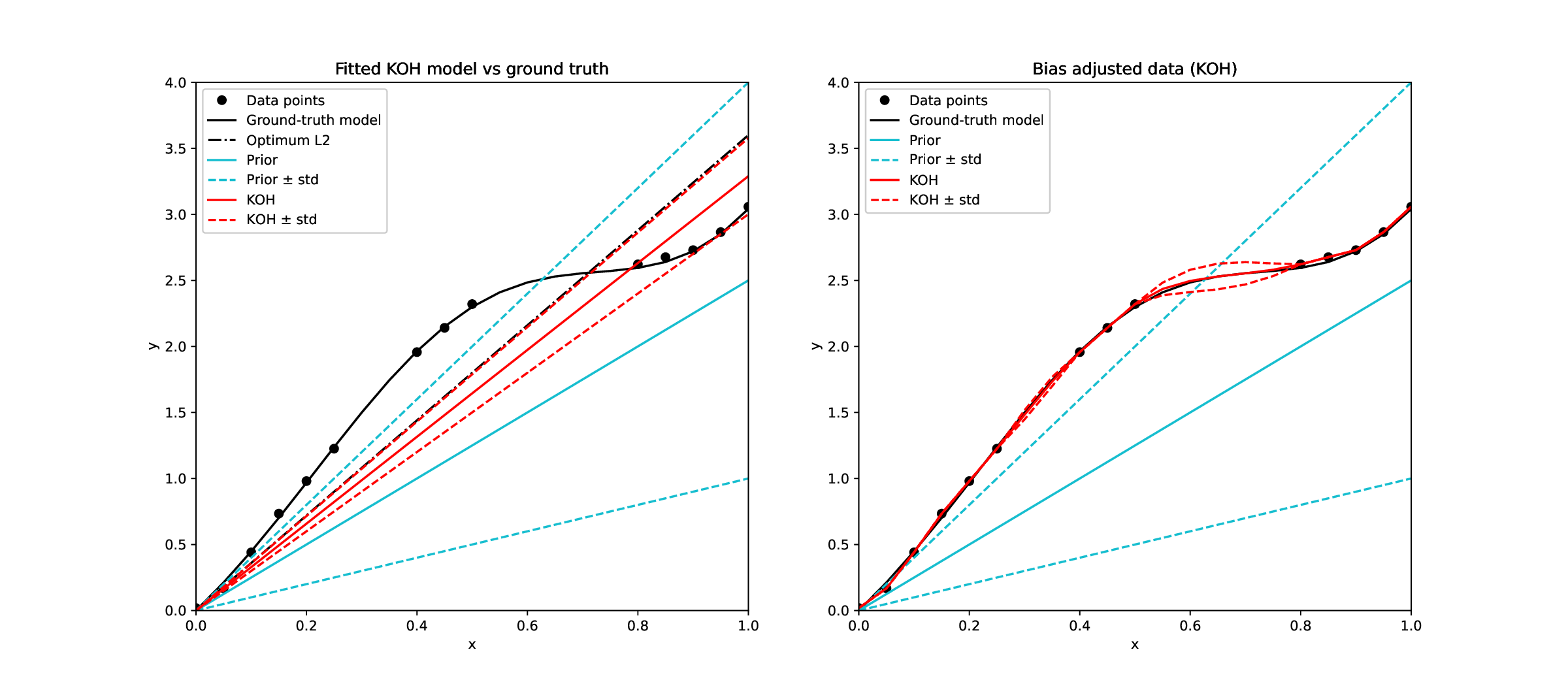}}
	\caption{System's response for Kennedy and O'Hagan's approach, 1D case. Fitted system with inferred $\theta$ (left) and bias-corrected response (right).}
	\label{fig:1d_koh_fitted}
\end{figure}

Next, the modularized version of KOH's approach presented in Algorithm \ref{alg:bayarri_kennedy_ohagan_gp} is applied to the benchmark case. A GP with a Matérn kernel of $\nu=3/2$ is used to model the bias term $b$. The correlation length for the bias GP is fixed to $\ell_b=\frac{0.5}{\sqrt{3}}$ in the interest of a representative comparison between methods. In this case, the input is one-dimensional and requires only samples of $x$ for training, as the bias is defined for a given $\theta$ and will be newly trained at every MCMC evaluation. Consequently, the training set is generated at every sample of $\theta$ and is composed by the positions $\hat{\mathbf{x}}$ of the measurements as input, and their corresponding residuals $\hat{\mathbf{r}}$ for a given $\theta$ as output. Note that for this approach, the anchor points $\xi$ of the GP is always the input set $\hat{\mathbf{x}}$. The fitted parameters can be observed in Table \ref{tab:Simple_inference}, and its fitted and bias-corrected responses are represented in Figure \ref{fig:1d_koh_fitted}. The bias is a GP regressor, therefore for the case of just one measurement per observation point and given a sufficient kernel, the bias-corrected predicted response interpolates the observations. However, in the regions without measurements, the response presents a significantly larger bias.

As expected, one of the biggest caveats for this approach is the identifiability problem that arises between the latent parameter and the bias term. By adding a highly non-linear bias, the optimal solution for this simple 1D case would be the fitted parameter collapsing to a Dirac's delta distribution with a value of 3.6 for the $L^2$ loss and the bias term compensating only the difference between the observations and the linear model. However, this is not the case, as the posterior predictive does not collapse despite the convergence of the MCMC process. This is explained due to the lack of identifiability between latent parameters and bias, still present in the modularized version of KOH, as the same improvement in the likelihood from Equation \ref{eq:gp_loglike} can be achieved by modifying the parameter towards the optimum to minimize the residuals or towards values that improve the variance of the GP.

\begin{figure}[ht]
	\centerline{\includegraphics[width=\textwidth]{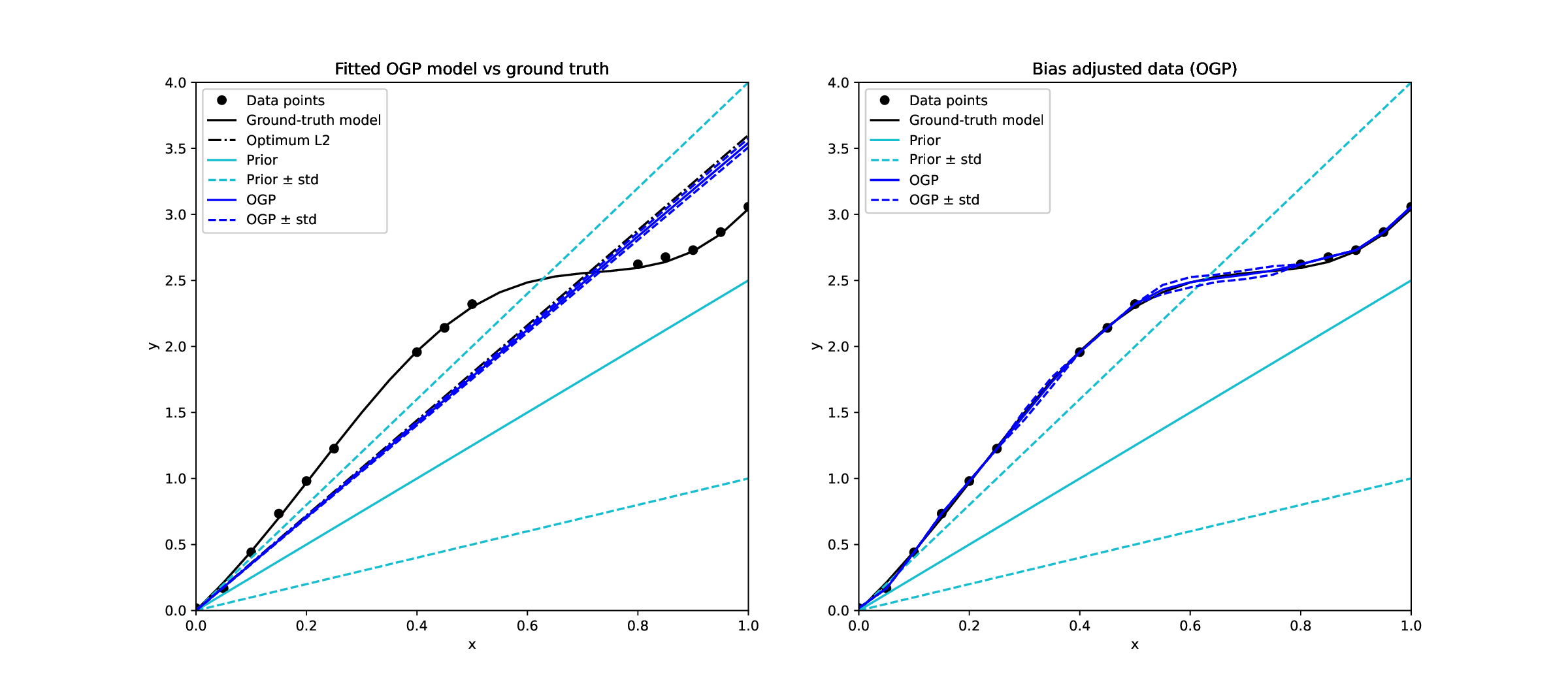}}
	\caption{System's response for Orthogonal GPs' approach, 1D case. Fitted system with inferred $\theta$ (left) and bias-corrected response (right).}
	\label{fig:1d_ogp_fitted}
\end{figure}

Analogously to the other approaches, OGPs are implemented for the simple 1D case. The main difference compared to KOH's is the imposition of the orthogonality condition between the bias and the latent parameters at every evaluation of the MCMC loop. The required derivatives are obtained by applying a simple finite differences method with a central scheme and an increment step $h=10^{-3}$. Additionally, the anchor points $\bm{\xi}$ are defined over the whole input domain $X$ discretized by the equidistant points $\mathbf{x}$ separated by intervals of 0.05. Consequently, the anchor points prescribe the orthogonality condition as well on the regions of the domain of interest without measurements. Every other parameter remains the same as for KOH, i.e. the prior distributions and correlation kernel. The inference results can be observed in Table \ref{tab:Simple_inference}, and Figure \ref{fig:1d_ogp_fitted} present the predictive results. The chosen metric is the $L^2$ loss, therefore the expected optimal value for $\theta$ is 3.6.

\begin{figure}[ht]
	\centerline{\includegraphics[width=\textwidth]{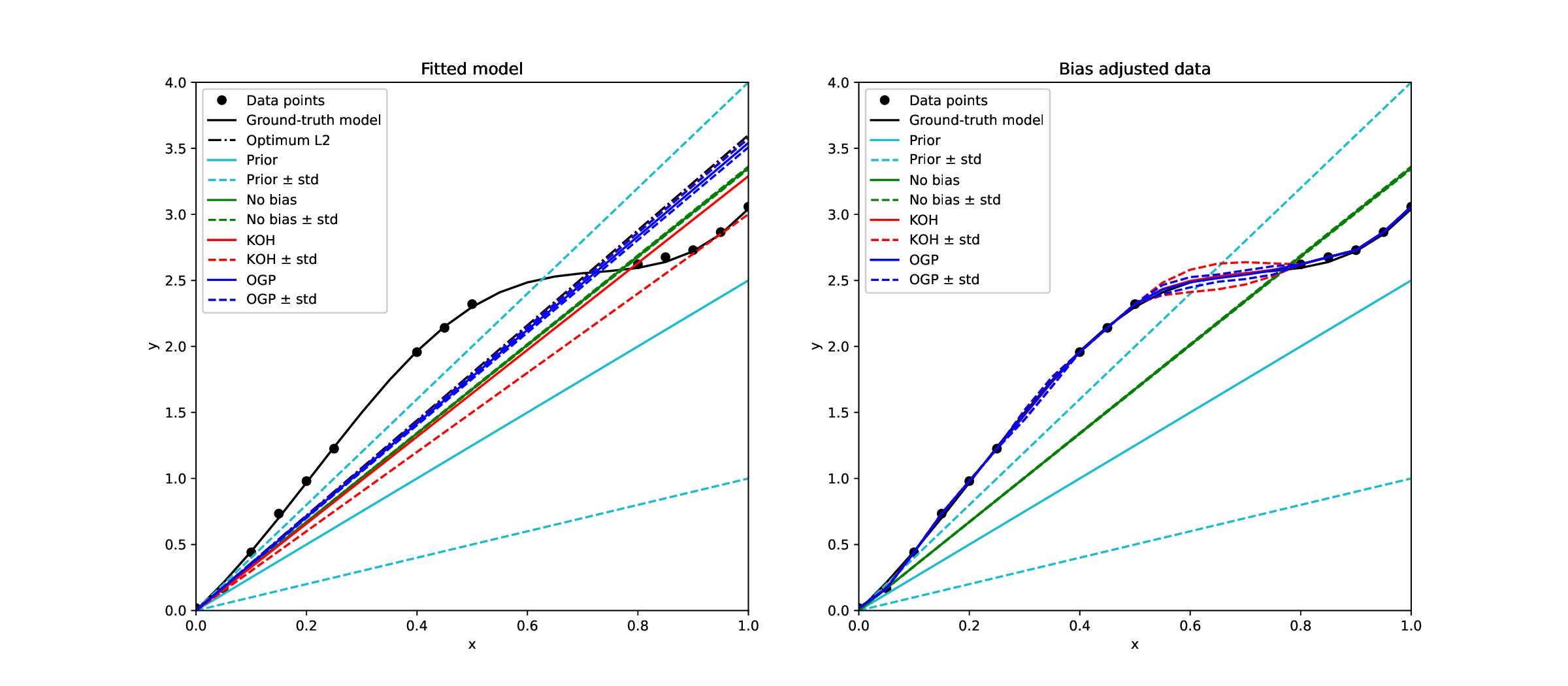}}
	\caption{Comparative system's response, 1D case. Fitted system with inferred $\theta$ (left) and bias-corrected response (right).}
	\label{fig:1d_comparative}
\end{figure}

A comparison of the posterior predictive results is represented in Figure \ref{fig:1d_comparative}. It can be observed that none of the fitted models covers the data points without considering the bias. As expected, the no-bias and OGP approaches converge to a Dirac's delta distribution, while KOH presents some variance. The analytical optimum value of $\theta$ using an $L^2$-norm without noise is 3.6, and for the Mean Square Error (MSE) on the predictions is 3.33. In the perfect case, the point estimate for the maximum likelihood of $\theta$ should converge to a chosen optimum with almost no variance in the parameter. Thanks to the definition of the anchor points $\hat{\bm{\xi}}$ over the whole domain, only OGP is able to achieve such precision and allows controlling the norm with respect to which the optimum is calculated.

Alternatively, when observing the bias-corrected prediction, all the methods that include the bias are able to cover the observations in the confidence region. KOH and OGP interpolate over the measurements and present a very low variance between them. This is expected, as they are best fitting GPs for the residuals between computational model and observations using a maximum likelihood estimator for $\theta$. 

Altogether, OGP presents the best behaviour at the expense of additional computation time. The running times for each approach on 4 CPUs of an AMD EPYC 74F3 processor and 8 GB of RAM are presented in Table \ref{tab:1d_times}. It can be observed that KOH and OGP multiply by more than 33 and 45 times respectively the computation time in comparison to not considering bias. For KOH and OGP, this increment comes from the training of the GP for the bias at every evaluation of the MCMC chain, with additional time in the case of OGP for the evaluation of the derivatives and setting the orthogonal prior. It is expected that, for more complex cases, the largest time is spent in evaluations of the model instead of in the training of the bias, whose computation time should not increase for KOH in comparison. For OGP, it would depend on the availability of the derivatives and the choice of anchor points, which would dictate the required additional model evaluations. In comparison, these approaches would be relatively more efficient for such complex cases. 

\begin{table}[ht]
	\centering
	\begin{tabular}{cc}
		\hline
		Method  & \multicolumn{1}{l}{Time (s)} \\ \hline
		No bias & 3.3                          \\
		KOH & 110.4 \\
		OGP  & 151.6 \\ \hline
	\end{tabular}
	\caption{Evaluation times for the first benchmark (simple case).}
	\label{tab:1d_times}
\end{table}

%%%%%%%%%%%%%%%%%%%%%%%%%%%%%%%%%%%%%%%%%%%%%%%%%%%%%%%%%%%%%%%%%%%%%%%%%%%%%%%%%%%%%%%%%%%%%%%%%%%%%%%%%
\subsection{Bias identification with heteroscedastic noise}
\label{sub:cantilever_beam}
This case aims to reflect the influence of model bias in the estimation of heteroscedastic noise. Two models are differentiated to this end: one simple cantilever beam under a deterministic load condition that is used as the prescribed computational model and another one biased and with a stochastic load used to generate the data. The objective is to infer the value of the Young's modulus for the computational model based on observations obtained from the biased one. This simplification mimics the situation of a testing load on a concrete cantilever bridge. The computational model supposes that only the experimental load is present in the real bridge, while the observations may have been collected with additional loads on the bridge or even with traffic. Additionally, it is supposed that a deficient placement and setting of the sensors leads to an additional displacement of $\Delta u=-100$ mm for every measurement. These extreme additions are introduced for illustrative purposes, but comparable effects are observed regularly in real implementations of bridge monitoring systems, especially if the model updating is done with data from the bridge in service. Table \ref{tab:beam_case} summarizes the case's formulation.

\begin{table}[htb]
	\centering
	\begin{tabular}{rp{0.6\linewidth}}
		\hline
		Element & Formulation \\ \hline
		Generator model  assumptions& Linear elastic cantilever beam that depends on the Young's modulus $E$ and the loads vector $\bm{\phi}$, and outputs the vertical displacements vector $\bm{u}$ at $\bm{x}$. $\Delta u(x=0\text{ m})=-0.1\text{ m}$, $\phi(x=50\text{ m})=100\text{ kN}+\delta_\phi$ with $\delta_\phi\sim\mathcal{LN}(20 \text{ kN}, 6\text{ kN})$.\\ 
		Generator model & $u^g(x,E)=g(x,E)+\epsilon$ for a given realization of $\delta_\phi$\\
		Computational model assumptions & Linear elastic cantilever beam that depends on the Young's modulus $E$ and the loads vector $\bm{\phi}$, and outputs the vertical displacements vector $\bm{u}$ at $\bm{x}$. $\Delta u(x=0\text{ m})=0.0\text{ m}$, $\phi(x=50\text{ m})=100\text{ kN}$.\\
		Computational model without bias term &$u^f(x,E)=f(x,E)+\epsilon$\\
		Computational model with bias term & $u^h(x, E)=u^f(x, \theta)+b(x)=f(x,E)+b(x)+\epsilon$ \\
		Discrepancy source & $\Delta u,~\delta_\phi$\\
		Latent parameter and prior & $E,~p(E)\sim\mathcal{N}(35 \text{ GPa}, 5 \text{ GPa})$, (with no bias, additionally $\sigma_\epsilon,~p(\sigma_\epsilon)\sim\mathcal{U}(0,0.8)$)\\
		Domain of interest  & $x\in X=[10\text{ m},50\text{ m}]$ \\ 
		Measurements' domain &$\hat{x}\in \hat{X}=[10\text{ m},50\text{ m}]$, observations at $\hat{x}=[10\text{ m},20\text{ m},30\text{ m},40\text{ m},50\text{ m}]$\\
		Noise & $\epsilon\sim\mathcal{N}(0, \sqrt{2\times10^{-8}})$\\
		\hline
	\end{tabular}
	\caption{Case description for Section \ref{sub:cantilever_beam} (beam model). The terms $f(x,E)$ and $g(x,E)$ represent the output of the FE model of the system under the assumptions of the computational model and the generator, respectively.}
	\label{tab:beam_case}
\end{table}

\begin{figure}[ht]
	\centerline{\includegraphics[width=0.9\textwidth]{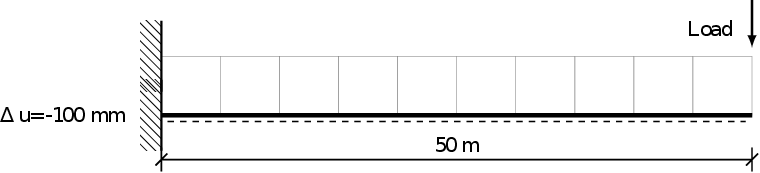}}
	\caption{Beam representation of the simplified system.}
	\label{fig:beam_schematic}
\end{figure}

Both models correspond to a Finite Elements (FE) \cite{Felippa2004} representation of such a bridge as a 2D cantilever beam of length $L=50$ m and height $H=3$ m, discretized with $30\times10$ quadrilateral linear elements. This represents half of the span of a bridge with constant a cross-section. The measurements will predict the deformation of the beam under a load at the free end with a simple linear elastic material law. The load represents a typical static test truck of 100 kN. The assumed width of the bridge is $W= 3$ and its Poisson's ratio $\nu$ is 0.2. Young's modulus $E$ is the subject of this study, but in general, it will have a base value of 30 GPa. A diagram for the model is represented in Figure \ref{fig:beam_schematic}.

The complex generative model includes the additional prescribed displacement at the fixed boundary, simulating a displacement at the abutment or a deficient setting of the sensors. Additionally, in contrast with the deterministic load for the computational model of 100 kN, the generative model adds a stochastic component. The result is a constant force of 100 kN plus a random draw from a log-normal distribution with mean 20 kN and standard deviation 6 kN. This variability represents a simplification of the additional load from a vehicle over the bridge. The origin of coordinates is located at the bottom part of the fix end of the beam. Vertical displacement measurements are obtained at $n=5$ virtual sensor positions situated at $\hat{y}=1.5$ m and $\hat{x}=[10\text{ m},20\text{ m},30\text{ m},40\text{ m},50\text{ m}]$, which correspond to points in the middle line of the beam. The simulation is repeated 20 times, each with a different realization of the load, generating as many measurements per observation point. A prescribed Gaussian noise of zero mean and standard deviation of $2\times10^{-8}$ m is added to the measurements to simulate typical sensor noise.

In this case, the virtual sensors cover sufficiently well the observation domain, therefore the fitted latent parameters are similar with and without bias. However, two impactful sources of error have been introduced: the prescribed displacement at the boundary and the variation in the load. They create respectively a discrepancy in the model's response compared to the measurements and additional noise with physical meaning that cannot be explained a priori. It is not possible to fully represent these effects without considering a bias term. First, the aforementioned discrepancy can be directly estimated by including the bias analogously as in the simple 1D case: adding it to the output as a GP with the same combination of constant and Matérn kernel $k_{GP}$. Second, to allow the same GP to consider multiple measurements per sensor coordinate, a noise-enabled kernel must be employed. Due to the noise coming from the variation in the load, its amplitude must reflect the physical behaviour of the beam, therefore the noise is expected to vary in scale at the different sensors. This requires employing a heteroscedastic noise component added to the kernel, whose parameters are trained during the phase where the bias is fitted. Such a component is using the heteroscedastic kernel implemented in \texttt{gp\_extras}\cite{Metzen2016}. A fixed homoscedastic white noise component with a $\sigma_N$ value of 1e-12 m is added to avoid numerical inconsistencies due to the proximity to zero noise at the sensors near the cantilever boundary. The resulting kernel follows
\begin{equation}
	k(x_i,x_j)=k_{GP}(x_i,x_j)k_{WN}(x_i,x_j)k_{HN}(x_i,x_j),
\end{equation}
where $k_{WN}(x_i,x_j)$ is defined as in Equation \ref{eq:homo_wn} and $k_{HN}(x_i,x_j)$ as in Equation \ref{eq:hetero_hn} where $\bm{\xi}=\hat{\mathbf{x}}$ and the $\sigma_i$ are fitted during the model evaluation.

Effectively, both the homoscedastic and heteroscedastic parts of the kernel will represent the same noise in the displacements produced by the load variability, which is expected to be larger than 1e-12 m at any sensor. Therefore, prescribing a constant homoscedastic part does not substantially impact the results, while increasing the numerical stability of the inference procedure. The additional derivatives required for OGP's approach will be calculated using a central finite difference scheme with a step of $h=1$ GPa. An alternative popular approach for estimating this noise term in the absence of bias is to include it as a parameter to be inferred. This approach increments the number of latent parameters, does not address the model discrepancy and requires prior information on the noise that is not generally available. Nevertheless, it will be applied in the case with no bias in the interest of a more detailed comparison, providing further noise estimation capabilities. 

\begin{figure}[ht]
	\centerline{\includegraphics[width=0.7\textwidth]{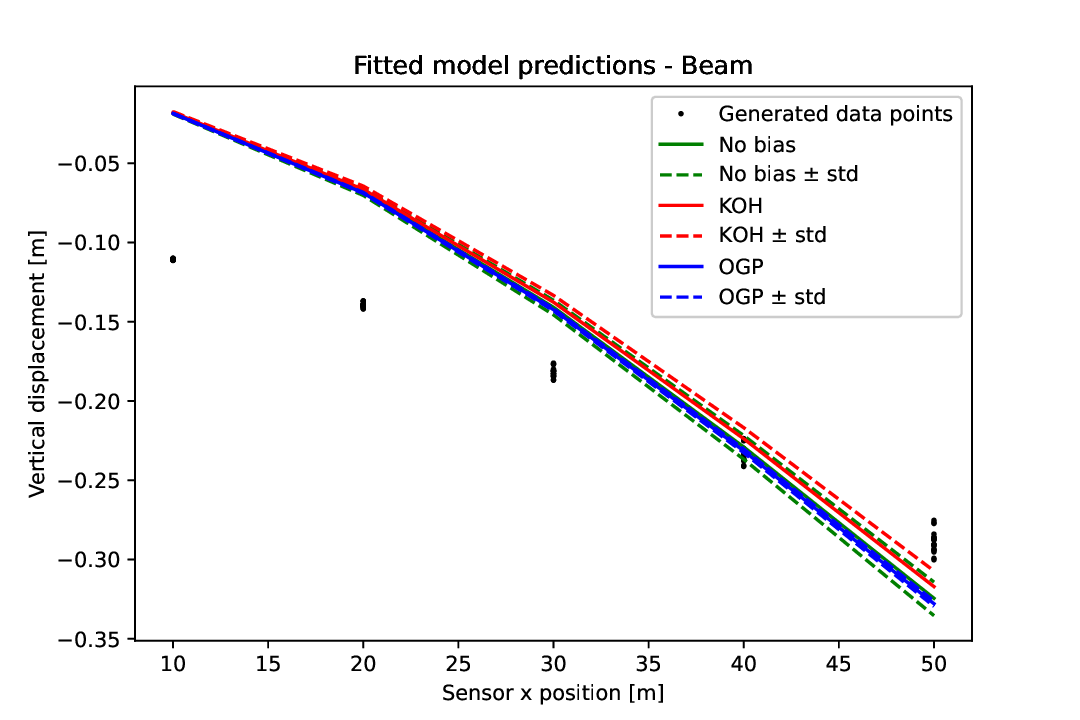}}
	\caption{Fitted system's response for the cantilever beam benchmark case with heteroscedastic noise.}
	\label{fig:beam_fitted}
\end{figure}

\begin{figure}[ht]
	\centerline{\includegraphics[width=0.7\textwidth]{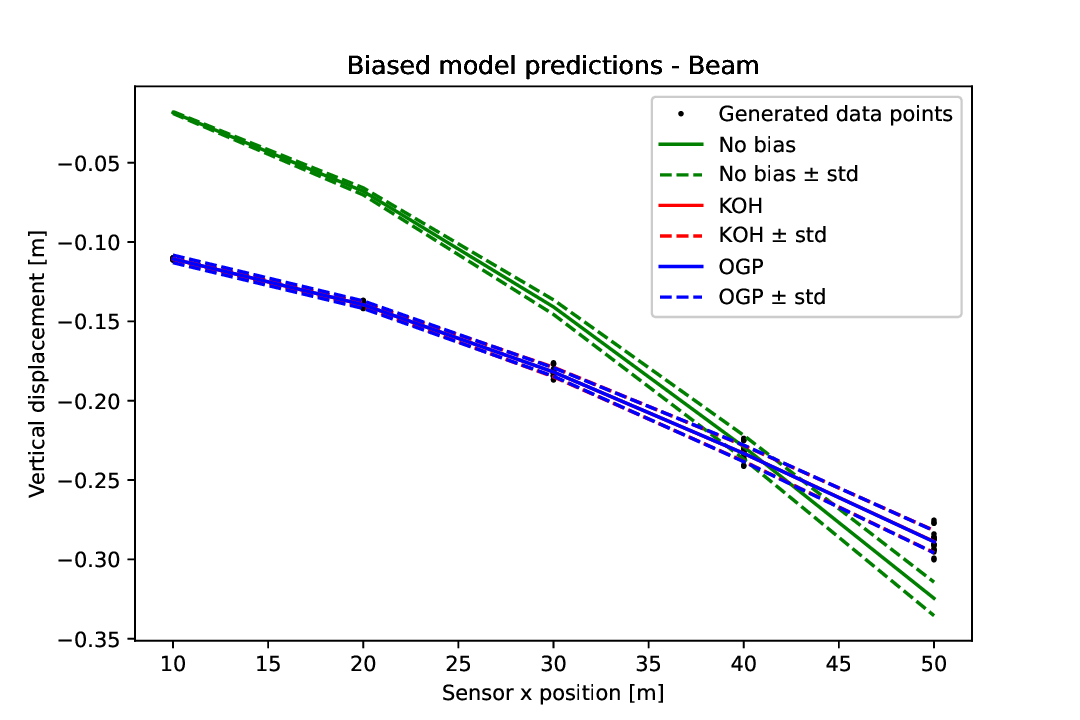}}
	\caption{Bias-corrected response for the cantilever beam benchmark case with heteroscedastic noise.}
	\label{fig:beam_biased}
\end{figure}

An MCMC algorithm is used to infer $E$ following the methodologies previously. The latent variable $E$ is modelled with a prior normal distribution $\mathcal{N}\sim(35\text{ GPa}, 5\text{ GPa})$. In the case without bias, the noise term $\sigma$ is defined with a uniform prior between 0 and 0.8 to allow any possible range of values. The MCMC algorithm is run for 500 steps with 100 additional steps of burn-in. The results can be observed in Figures \ref{fig:beam_fitted} and \ref{fig:beam_biased}. The summary statistics are displayed in Table \ref{tab:beam_inference}.

\begin{center}
\begin{table*}[!h]
	\centering
\begin{tabular}{llccccccc}
\toprule
  Group & Parameter &     Mean &        SD &    HDI$_{3\%}$ &   HDI$_{97\%}$ &  MCSE$_{\text{mean}}$ &   MCSE$_{\text{SD}}$ &     $\hat{R}$ \\
\midrule
No bias &       $E$ & 1.486e+10 & 4.838e+08 & 1.391e+10 & 1.579e+10 &  3.999e+07 & 2.833e+07  & 1.070e+00 \\
No bias &  $\sigma_\epsilon$ & 5.900e-02 & 5.000e-03 & 5.100e-02 & 6.800e-02 &  0.000e+00 & 0.000e+00 & 1.070e+00 \\
KOH &       $E$ & 1.521e+10 & 4.907e+08 & 1.428e+10 & 1.609e+10 &  3.609e+07 & 2.561e+07 & 1.040e+00 \\
OGP &       $E$ & 1.470e+10 & 7.393e+07 & 1.459e+10 & 1.483e+10 &  1.177e+07 & 8.384e+06 & 1.170e+00 \\
\bottomrule
\end{tabular}
\caption{Summary inference results for the beam case. Mean and SD are the statistics of the sampled parameters. HDI 3\% and 97\% are the bounds of the 94\% Highest Density Interval. MCSE$_\text{mean}$ and MCSE$_\text{SD}$ are the Monte Carlo Standard Errors due to having a finite number of samples in the MCMC for the mean and standard deviation, respectively. $\hat{R}$ is the Gelman-Rubin diagnostic\cite{Gelman1992} (1 = perfect convergence).}
\label{tab:beam_inference}
\end{table*}
\end{center}


% Comments on no-bias
Convergence is achieved with every method in the chosen number of steps. For the approach without bias, the noise is inferred as an extra latent parameter. It must be noted $E$ and $\sigma_\epsilon$ do not allow for a unique identification of such parameters due to the effect of the model discrepancy. Nevertheless, the convergence is apparent. Independently of the scale of the Gaussian noise $\sigma$, the fitted model will not be able to replicate the generated dataset. In comparison with the real Young's modulus with a mean of 30 GPa and a standard deviation of 10 GPa, the inference procedure converges to a mean of around 14.9 GPa and standard deviation 0.5 GPa, which means an error of more than 50\%. This relatively large difference is expected, as the predictions will never be able to accommodate the observations due to the prescribed displacement in the generative process. To obtain predictions closer to the observations, the computational model tends to become less stiff to accommodate the offset in the observations. The magnitude of the inferred noise without bias remarkably ranges between 0.051 and 0.068 m. With these values, it would fail to represent the noise both near the fixed end, close to zero, and at the free end, of approximately 0.025 m for the sampled data. Therefore, these predictions from the fitted model do not guarantee that the computational model reflects the real system, which is a requirement in the implementation of digital twins.

% Comments on KOH
Similarly, the method that adds a bias term following KOH's approach converges to a fitted Young's modulus of mean 15.2 GPa and a standard deviation of 0.5 GPa. Once again, the system converges to the optimum fitting value that compensates for the deviation, not being able to represent the real generative process. The bias correction is consequently applied to the maximum-likelihood estimator of the fitted model, which generates corrected predictions. The key difference is the introduction of the heteroscedastic noise kernel, which allows capturing the variability in the observations.

% Comments on OGP
Alternatively, adding the bias term as an OGP leads to a slower convergence measured by the Gelman-Rubin diagnostic. Imposing the orthogonality condition in the kernel with a noise component leads to GP kernel proposals that may not be able to fit the observations efficiently for some samples. To avoid this effect, the non-orthogonal part of the kernel could be adapted with a larger correlation length and scale, instead of using the same as in KOH's approach. In the interest of an equivalent comparison, the chosen kernel prior was the same for both approaches. Despite this slightly larger $\hat{R}$ of 1.17 after 500 MCMC steps, the OGP approach reasonably converges to a value of 14.7 GPa with a standard deviation of around 0.07 GPa, analogously to the other approaches in mean but with a decreased uncertainty. The noise structure and bias correction are equivalent to that presented in KOH's approach. Nevertheless, the MCMC algorithm takes more than 5 times longer to evaluate due to an inefficient differentiation scheme and the introduction of poorly behaved priors for samples with extreme values. Due to the heteroscedastic noise kernel, which introduces a noise hyperparameter for each of the sampled positions, the training of the bias OGP may fail to converge more often when the strict orthogonality condition prevents finding the extended set of optimum noise hyperparameters. This could be improved by relaxing the orthogonality condition, which is possible by adjusting the fixed parameters of the Matérn kernel to increase the reach of the non-orthogonal part of the GP.

% Comments on comparative
In comparison, the three approaches converge to a similarly wrong value of Young's modulus in the fitted model due to the inability of the computational system to reflect the observations with the additional deformation. The optimal value for $E$ such that either the MSE or the $L^2$ loss between observations and predictions are minimized is around 14.8 GPa, which is approximately reached by each of the approaches. As both MSE and $L^2$-optimum values are relatively close, it was expected that both the OGP and the unbiased method converge to a similar value. KOH's approach tends to those same values, as these optimum parameters would provide the best-fitting bias distribution for the chosen GP prior. Adding the noise as a tunable parameter in the approach without bias does not improve the result nor quantify the real noise at the output. Contrarily, adding the bias term as in either approach allows the correction of the bias and quantifies the noise reliably. In this case, there is no apparent benefit of using OGP over KOH despite its larger computational times (see Table \ref{tab:times_beam}), as the chosen prior for KOH leads to the same optimum.  As the bias term is directly applied at the output of the system, the value of $E$ does not reflect the one in the generative model and therefore could not be used for derived quantities that depend on it. 

\begin{table}[ht]
	\centering
	\begin{tabular}{cc}
		\hline
		Method  & Time \\ \hline
		No bias & 30 min                          \\
		KOH  & 4 h 50 min                        \\
		OGP  & 21 h 27 min \\ \hline
	\end{tabular}
	\caption{Evaluation times for the second benchmark (beam case).}
	\label{tab:times_beam}
\end{table}

The key feature highlighted by these results is the capacity of the bias term to represent a deviation of a biased model from reality. The noise in the approach without bias term cannot be quantified appropriately due to the existing discrepancy, while KOH can reliably correct the prescribed displacement and infer the noise produced by the variations in the material. This physical noise component of the observations is included as part of the model bias, which agrees with the assumption that the model's Young's modulus employed in the digital twin is unique and deterministic. The inability of the model to represent stochastic responses is compensated by a noise-enabled bias with a modified kernel. Without this bias term, the model response variance remains unknown and the model predictions are not able to represent the reality of the system. It is shown that a heteroscedastic noise kernel can be used in the bias to this end in a non-intrusive manner, without modifying the model or the inference schemes. Samples from this posterior predictive distribution can be used in the context of a digital twin to provide information on points in $X$ where no observation is available. The stochastic nature of the bias-corrected predictions provides a quantification of the uncertainty in the observations produced by the variability of the real system, and not only due to the indeterminacy of the model parameters. For example, large variations in the bias could indicate a deficient testing setup, or in this case, the inclusion of loads not considered in the model.

%%%%%%%%%%%%%%%%%%%%%%%%%%%%%%%%%%%%%%%%%%%%%%%%%%%%%%%%%%%%%%

\subsection{Bias identification incorporating additional environmental conditions}
\label{sub:bridge}
\subsubsection{Case definition}
\begin{figure}[ht]
	\centerline{\includegraphics[width=0.9\textwidth]{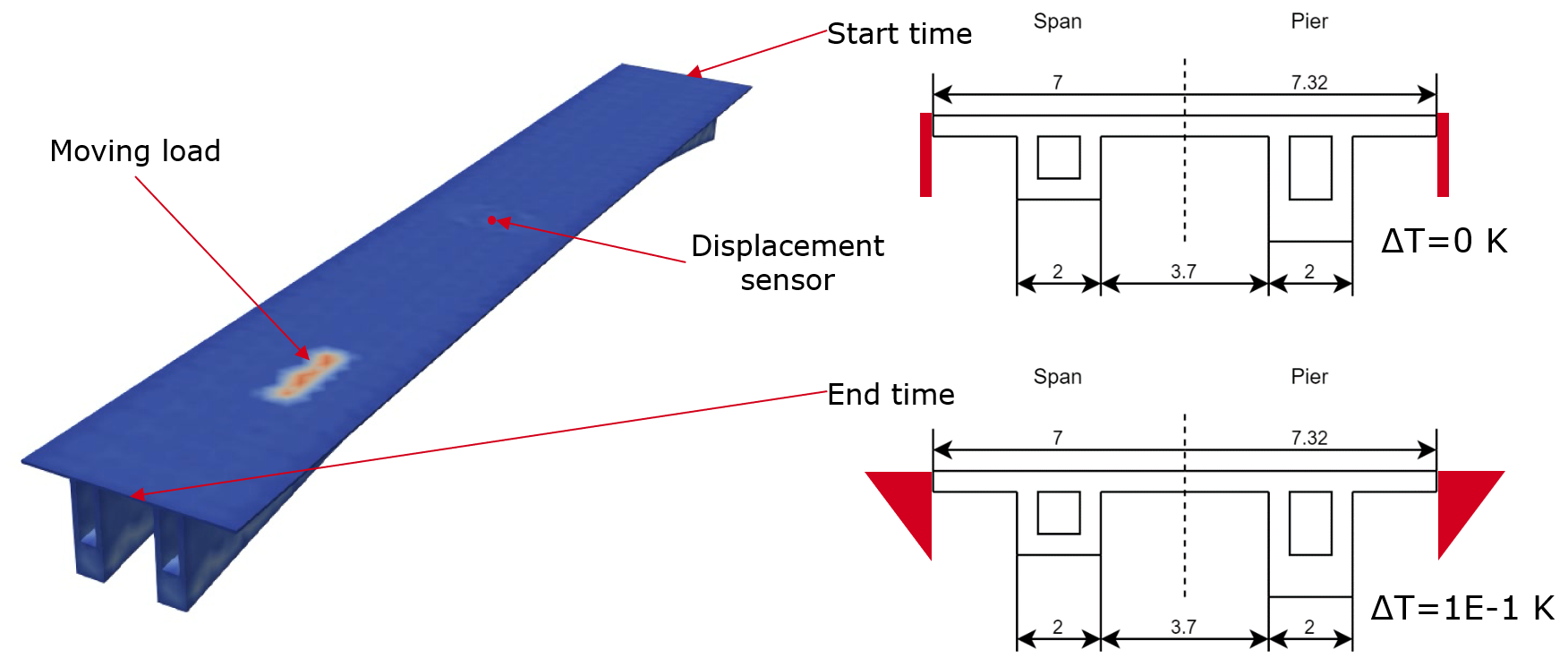}}
	\caption{Nibelungenbrücke experiment scheme. Measurements in m, not at scale.}
	\label{fig:nibelungenbruecke_model}
\end{figure}

The objective of this case is the inclusion of additional information on the environmental conditions in the bias identification applied to a real use case. Consequently, the final use case illustrates their implementation on a simplified model of an actual bridge: the Nibelungenbrücke in Worms, Germany, displayed in Figure \ref{fig:nibelungenbruecke_model}. The research initiative "Schwerpunktprogramm 2388: SPP Hundert Plus", within which this work is included, aims to monitor and eventually prolong the remaining life of the bridge through novel digital approaches, in particular the implementation of digital twins\cite{SPP2023}. Therefore, a model using a simplified version of Nibelungenbrücke's geometry is developed. The objective of this use case is the evaluation of the model bias approaches in a digital twin setting and the inclusion of additional variables, therefore the modelling of the bridge has been extremely simplified and would not necessarily coincide with an accurate representation of the real one. It must be particularly noted that the spans of the real bridge should be represented as cantilever structures attached at the center of the spans and with the pilots as supports. However, the simulation of a span will be represented by a simply-supported structure of the full span length for illustrative purposes. This choice has no impact on the conclusions extracted from this benchmark with respect to the application of biased approaches. The work presented in this paper will use simulation models for both generating the virtual observation data and predicting results. Measurements from the bridge are expected in the future, which will allow further testing of the methodology under real sensor observations. However, the validation with real data remains out of the scope of this paper, and this benchmark reflects the real system with many simplifying assumptions. Table \ref{tab:bridge_case} presents a summary of the formulation for this case.

\begin{figure}[ht]
	\centerline{\includegraphics[width=0.7\textwidth]{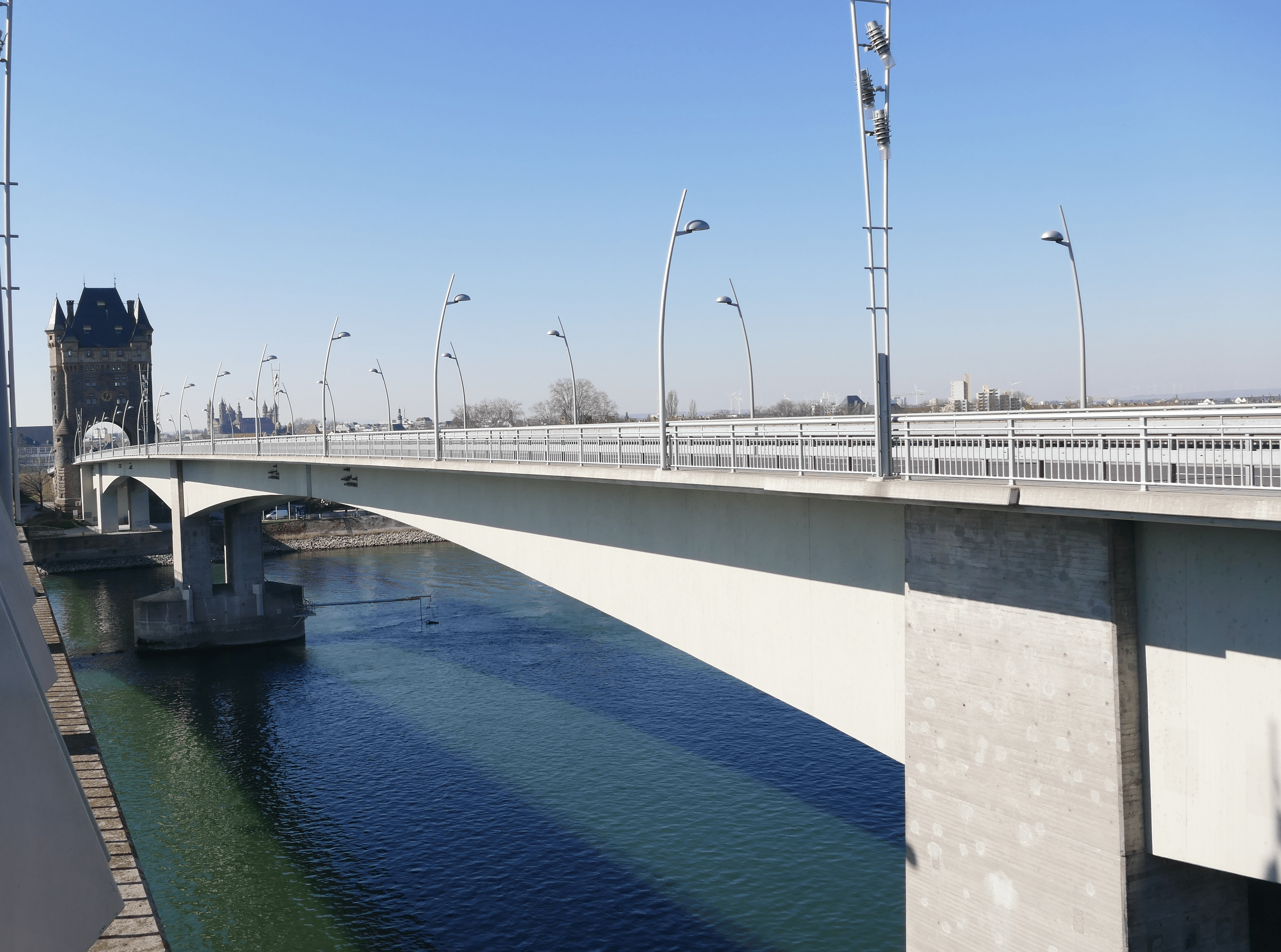}}
	\caption{Nibelungenbrücke photo. Source: Chongjie Kang, TU Dresden.}
	\label{fig:nibelungenbruecke_photo}
\end{figure}

\begin{table}[ht]
	\centering
	\begin{tabular}{rp{0.6\linewidth}}
		\hline
		Element & Formulation \\ \hline
		Generator model  assumptions& Load line-test of linear elastic simplified bridge model that depends on the Young's modulus $E$, the position $x$ of a known test truck load of 10 t and the temperature difference $\Delta T$ between start and end of the experiment, and outputs the vertical displacement $u$ at $x = 47$ m. \\ 
		Generator model & $u^g(x,\Delta T,E)=g(x, \Delta T,E)+\epsilon$\\
		Computational model assumptions & Load line-test of linear elastic simplified bridge model that depends on the Young's modulus $E$ and the position $x$ of a known test truck load of 10 t, and outputs the vertical displacement $u$ at $x = 47$ m.\\
		Computational model without bias term &$u^f(x,E)=f(x,E)+\epsilon$\\
		Computational model with bias term (without temperature) & $u^h(x, E)=u^f(x, \theta)+b(x)=f(x,E)+b(x)+\epsilon$ \\
		Computational model with bias term (with temperature) & $u^h(x, \Delta T, E)=u^f(x, \theta)+b(x, \Delta T)=f(x,E)+b(x, \Delta T)+\epsilon$ \\
		Discrepancy source & Additional $u$ induced by $\Delta T$\\
		Latent parameter and prior & $E,~p(E)\sim\mathcal{LN}(35.8 \text{ GPa}, 8.9 \text{ GPa})$\\
		Domain of interest  & Load position $x\in X=[0\text{ m},104\text{ m}]$ $\times$ $\Delta T\in \Theta=[0.01 \text{ K}, 0.1 \text{ K}]$ \\ 
		Measurements' domain &$(\hat{x},\hat{\Delta T})\in (\hat{X}\times\hat{\Theta})=\left( [0\text{ m},0.01 \text{ K}],[104\text{ m}, 0.1 \text{ K}]\right) $, observations at $\Delta T = $[0.01, 0.028, 0.046, 0.064, 0.082, 0.1] K for $x$ at intervals of 1 m.\\
		Noise & $\epsilon\sim\mathcal{N}(0, \sqrt{1\times10^{-12}})$\\
		\hline
	\end{tabular}
	\caption{Case description for Section \ref{sub:bridge} (bridge model). The terms $f(x,E)$ and $g(x,E)$ represent the output of the FE model of the system under the assumptions of the computational model and the generator, respectively.}
	\label{tab:bridge_case}
\end{table}

\begin{figure}[ht]
	\centerline{\includegraphics[width=0.9\textwidth]{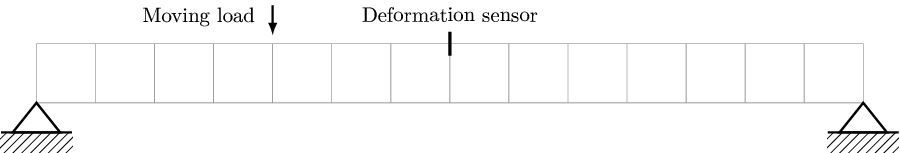}}
	\caption{Bridge simplified system diagram.}
	\label{fig:nibelungenbruecke_system_diagram}
\end{figure}

The first span of the bridge is chosen for the simulation model. It is represented as a solid 3D Finite Element (FE) box-girder model with variable cross-section. The dimensions of the model correspond to those indicated in available drawings\cite{Pelke2015} and databases\cite{Janberg2000}, with a span length of 95,185 m, span width varying between 14 m at the centre and 14.64 at the pilots, deck thickness of 0.25 m, cross-section height at pilots of 6.5 m and of 2.5 m at the middle of the span. The mesh is a tetrahedral 3D solid unstructured mesh with linear Lagrangian elements, with characteristic length between 0.2 and 2.0 m. The material is modelled using a purely linear elastic constitutive law with a Young's modulus of 40 GPa, a Poisson's ratio of 0.2 and a density of 2350kg/m$^3$. Simply supported boundary conditions are considered at the end surfaces, where the bottom edge of the supports is fixed in displacements, as represented in Figure \ref{fig:nibelungenbruecke_system_diagram}. The resulting model has 3530 nodes, which is equivalent to 10580 Degrees of Freedom (DoFs) and 20669 elements. A more complex model is possible, but it would not enhance in any way the conclusions from this study.
The virtual experiment will represent an influence line test under a moving load. A virtual displacement sensor is placed at the middle point of the span on the surface of the deck to measure vertical displacements. Then a simulated moving load is applied passing over the bridge, and the measurements of the sensor are recorded for every load position. In this case, the load represents a truck of 10 t, 8 m long and 3 m wide moving at a speed of 0.5 m/s over the bridge. This load is modelled as a rectangular surface load of the same dimensions, taking the front line of the truck as the reference for its position. The displacements are recorded with a sampling rate of 2 s. The plot of the observations recorded by the sensor during the time the truck is passing over the bridge is considered the influence line of that load. This line can be used to infer parameters of the bridge such as its Young's modulus. The calculation of the influence load does not consider any inertial effects, hence the simulation is a purely static one evaluated with different load locations.

\begin{figure}[ht]
	\centerline{\includegraphics[width=0.7\textwidth]{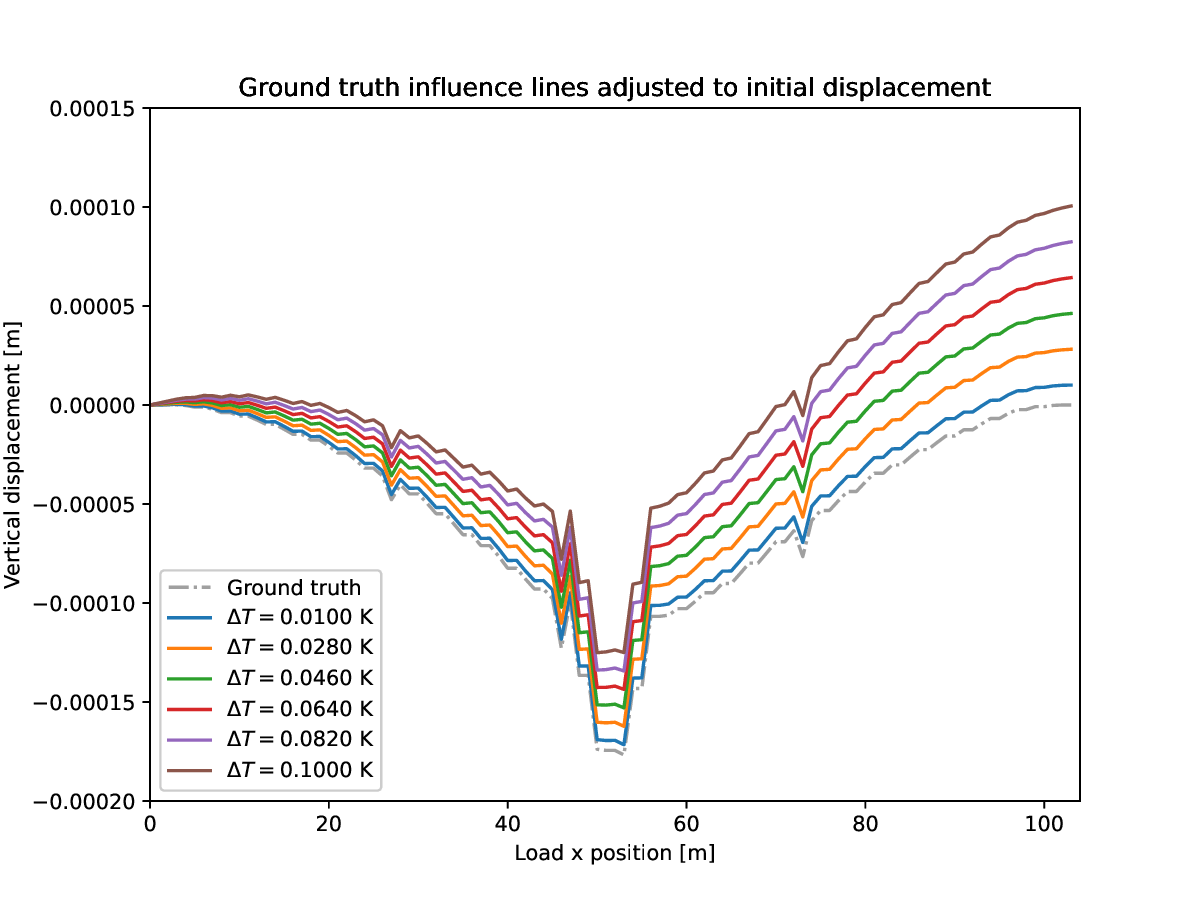}}
	\caption{Influence lines for the bridge case.}
	\label{fig:bridge_influence_lines}
\end{figure}

The computational model to be tuned will be the simulation to generate the influence lines and the parameter of interest that must be inferred will be the Young's modulus used in the simulation. An additional temperature effect is added only for the generation of the dataset, introducing a discrepancy between the generated data and the computational model. During the time that the truck would pass over the bridge, it is reasonable to assume a change in the temperature of the top surface of the deck\cite{Wang2021}. This can be, for example, due to the incidence of the sun over it. At the start of the experiment, we support a constant stationary temperature over the whole bridge. The top part of the deck will increase its temperature due to the transient effect of the heat on the surface, while the bottom part will remain unaffected due to the short timescale that prevents fast heat propagation. This situation generates a linear temperature gradient along the cross-section of the deck, which implies further thermal strains that will be identified by the sensor. These strains are introduced in the linear elastic model with a thermal expansion coefficient $\alpha=10^{-5}$.  A diagram of the experimental set-up is represented in Figure \ref{fig:nibelungenbruecke_model}. In practice, the difference in temperature between the start and end of the measurements on the top of the deck is measured, supposing a linear evolution of it. In this case, we run 6 simulations with temperature differences $\Delta T$ between start and end of [0.01, 0.028, 0.046, 0.064, 0.082, 0.1] K. This corresponds to a change of between 5 and 15 K in a period of 12 h, which is in line with expected values in similar applications\cite{Xia2018}. The observed influence lines for this experiment are represented in Figure \ref{fig:bridge_influence_lines}. The base case with no temperature difference has been represented in the figure as a reference, but will not be fed as data to the model for its calibration. Notably, the deformation is greater than zero at the end of the simulation when the temperature gradient is introduced. This effect cannot be replicated by the original model, as the only external load applied on the system comes from the passing truck, which at the end of the measurements is no longer over the bridge. Therefore, all the fitted models will predict zero deformation at the end of the simulation. No additional noise is introduced in the system except for a negligible value added for numerical stability, although the same methodology of the previous benchmark could be applied.

\subsubsection{Models without temperature consideration}
Analogous to the previous benchmark cases, an MCMC algorithm is used to infer Young's modulus $E$ from its influence lines. In this case, $E$ is prescribed with a prior log-normal distribution with parameters $\log\mu=24.3$ and $\log\sigma=0.2$. The equivalent mean for $E$ is approximately 35.8 GPa, which lies within a reasonable range of the real value of 40 GPa. The MCMC algorithm is run for 1000 steps with 200 steps of burn-in. For the case without bias, the 6 influence lines are treated as independent datasets, each of them weighted equally in the likelihood calculation. For KOH and OGP, an additive bias term modelled as a GP is included in the computational model. First, the bias GPs will be implemented analogously as the previous cases, without extending the input domain, following the algorithms from Section \ref{sec:model_bias_approaches}. In the case of KOH's approach, the kernel of the bias GP is a Matérn one with a fixed lengthscale $\ell=\frac{5}{\sqrt{3}}$ and a fixed standard deviation $\sigma=1.0$, multiplied by a constant valued kernel that is variable. For OGP, this same kernel is applied as the base to which the orthogonality condition is imposed. The load position for each observation is used as training input for the bias, and the residual of a given observation for the predictions is taken as output. Notice that there are 6 distinct observations for each load position with their corresponding residuals, one for each experimental series. These observations are indistinguishable from each other for the model and the GP, as they do not contemplate changes in the temperature. The predicted bias correction for a given load position will therefore be unique, independently of the temperature difference at the end of the experiment. 

\begin{center}
\begin{table*}[!h]
	\centering
\begin{tabular}{lccccccc}
\toprule
Group &     Mean &        SD &    HDI$_{3\%}$ &   HDI$_{97\%}$ &  MCSE$_{\text{mean}}$ &   MCSE$_{\text{SD}}$ &     $\hat{R}$ \\
\midrule
                No bias & 5.785e+10 & 1.952e+05 & 5.785e+10 & 5.785e+10 &  2.105e+04 & 1.494e+04 & 2.950e+00 \\
KOH without Temperature & 3.998e+10 & 1.698e+08 & 3.966e+10 & 4.026e+10 &  1.699e+07 & 1.206e+07 & 1.080e+00 \\
KOH with Temperature & 4.000e+10 & 4.321e+06 & 3.999e+10 & 4.001e+10 &  3.596e+05 & 2.548e+05 & 1.050e+00 \\
OGP without Temperature  & 5.783e+10 & 6.487e+07 & 5.772e+10 & 5.797e+10 &  1.506e+07 & 1.082e+07 & 1.240e+00 \\
OGP with Temperature  & 5.778e+10 & 3.703e+08 & 5.778e+10 & 5.800e+10 &  5.823e+07 & 4.169e+07 & 1.580e+00 \\
\bottomrule
\end{tabular}
\caption{Summary inference results for the parameter $E$ in the bridge case. Mean and SD are the statistics of the sampled parameters. HDI 3\% and 97\% are the bounds of the 94\% Highest Density Interval. MCSE$_\text{mean}$ and MCSE$_\text{SD}$ are the Monte Carlo Standard Errors due to having a finite number of samples in the MCMC for the mean and standard deviation, respectively. $\hat{R}$ is the Gelman-Rubin diagnostic\cite{Gelman1992} (1 = perfect convergence).}
\label{tab:bridge_inference}
\end{table*}
\end{center}

\begin{figure}[ht]
	\centerline{\includegraphics[width=0.7\textwidth]{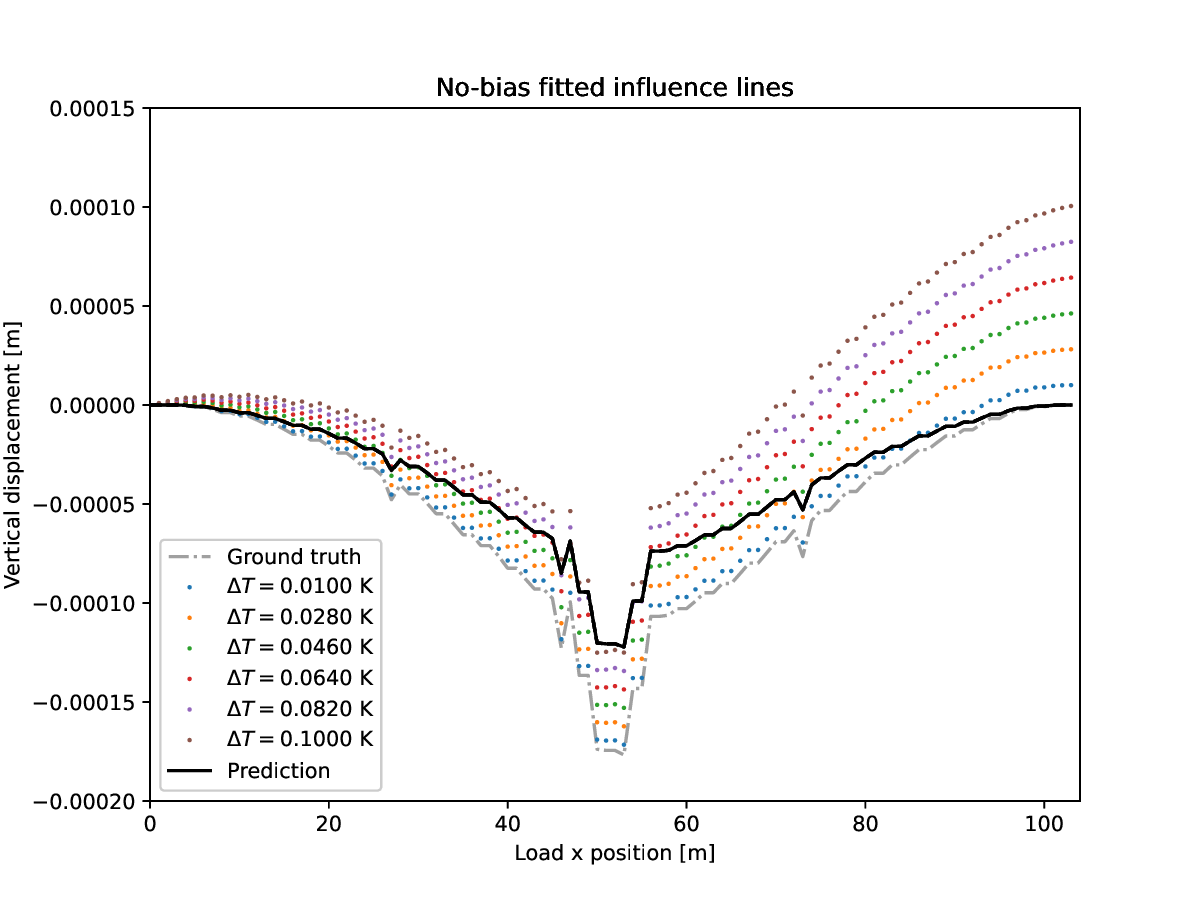}}
	\caption{Fitted system's influence lines obtained with the approach without bias. The points represent the generated observations. Each line represents the predicted response for one experiment, which coincide for this case.}
	\label{fig:bridge_nobias}
\end{figure}

% No-bias approach
Inference results can be observed in Table \ref{tab:bridge_inference}. Due to the absence of a non-prescribed noise latent parameter, the approach without bias tends to minimize the MSE for each load position. Therefore, it converges to a value of around 57.8 GPa with strong certainty, which implies an error of 44.5\% with respect to the real value. The resulting influence lines of fitting the model without including a bias term are represented in Figure \ref{fig:bridge_nobias}. The fitted model is not representative of the real system despite its certainty. This situation renders the obtained results unreliable and inaccurate. If a noise term were to be included, still the system would not reflect the reality, as the inferred parameters would not coincide with the real ones, and the generated predictions could not be used for making decisions on the behaviour of the bridge. In particular, the non-zero response at the end of the experiment despite the absence of load on the system cannot be replicated without considering a bias.

\begin{figure}[ht]
	\centerline{\includegraphics[width=\textwidth]{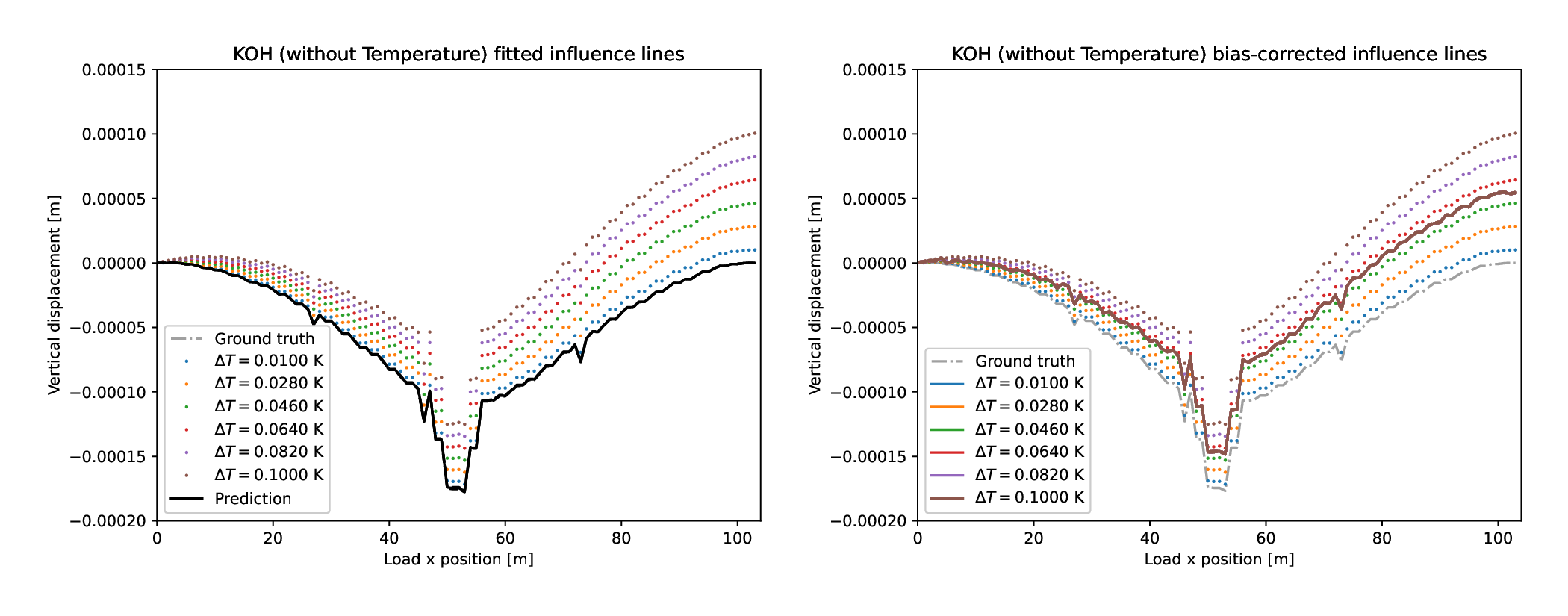}}
	\caption{Predicted response of the fitted model using KOH's approach without including the temperature information. Left, fitted response of the computational model. The prediction coincides almost exactly with the ground truth, and its variance is not visible. Right, bias-corrected response for each temperature series. All the response lines coincide in the plot.}
	\label{fig:bridge_koh_no_temp}
\end{figure}

\begin{figure}[ht]
	\centerline{\includegraphics[width=\textwidth]{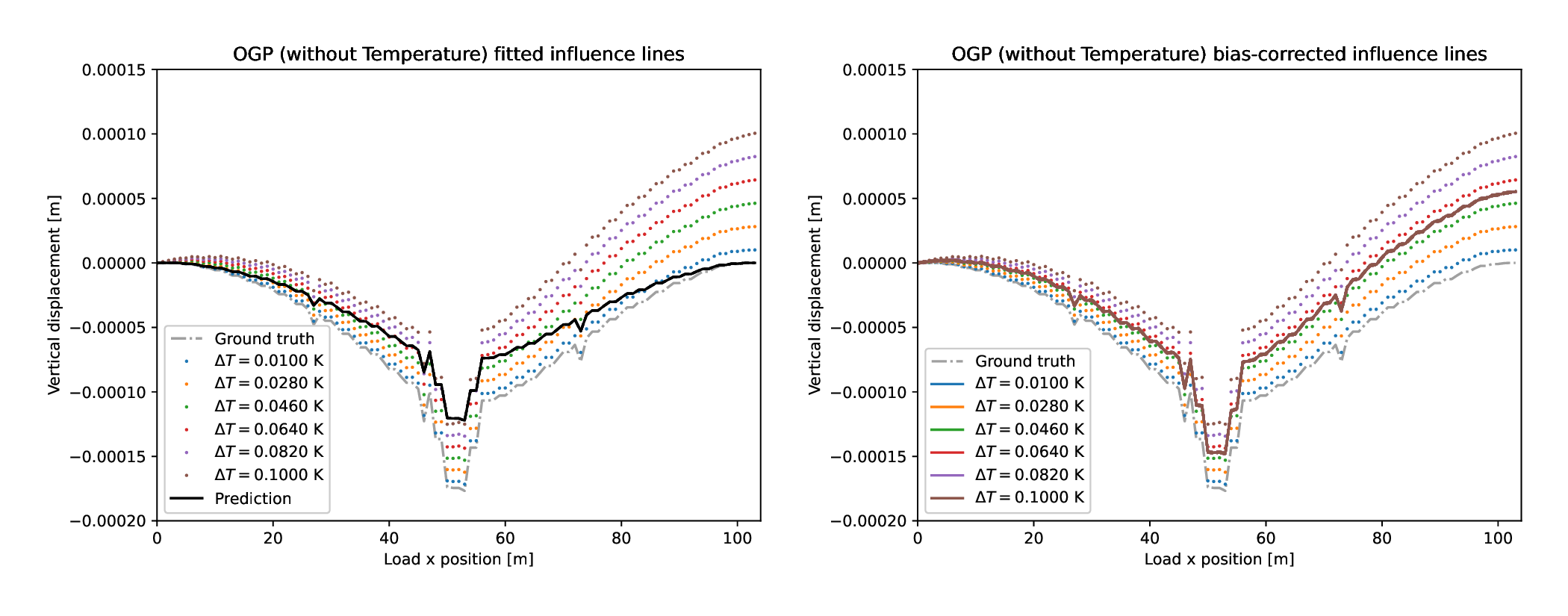}}
	\caption{Predicted response of the fitted model using OGPs without including the temperature information. Left, fitted response of the computational model. The prediction's variance is not visible. Right, bias-corrected response for each temperature series. All the response lines coincide in the plot.}
	\label{fig:bridge_ogp_no_temp}
\end{figure}

\begin{figure}[ht]
	\centerline{\includegraphics[width=\textwidth]{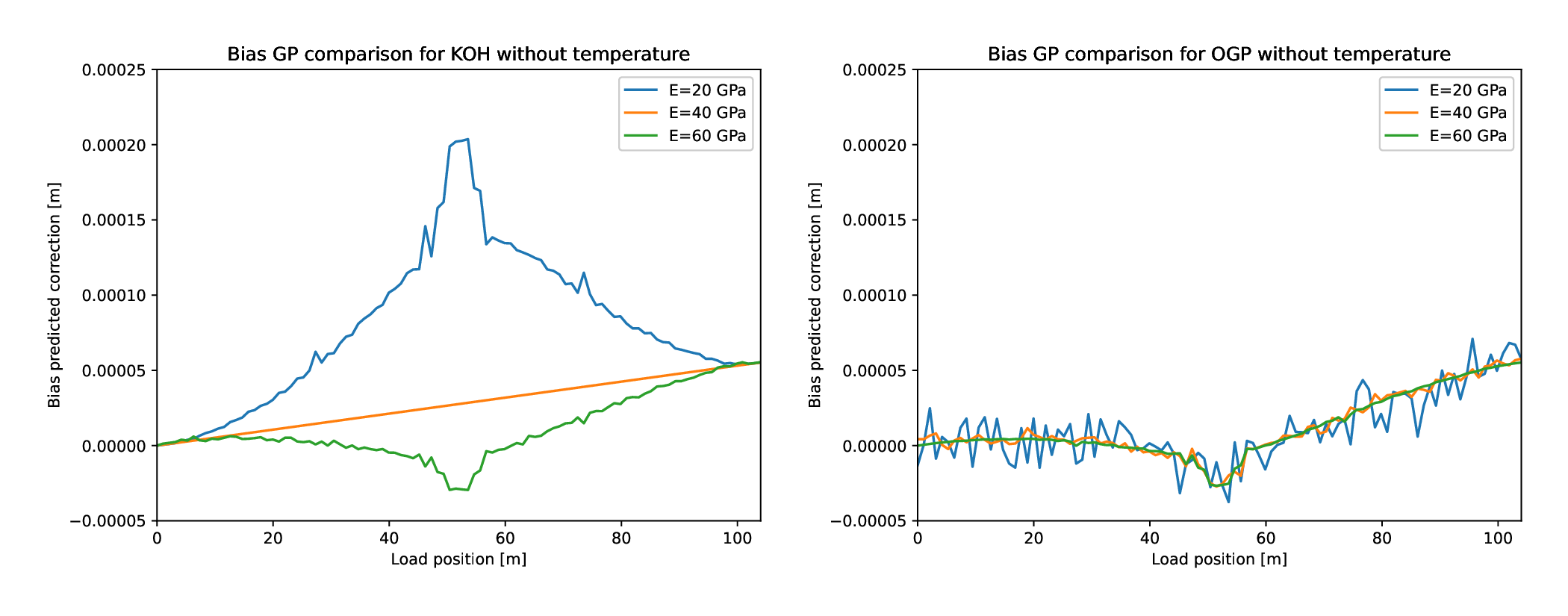}}
	\caption{Comparison of bias corrections for $E$=[20, 40, 60] GPa without temperature awareness. KOH's approach (left) and OGPs approach (right).}
	\label{fig:bridge_bias_no_temp_comparison}
\end{figure}

Alternatively, both KOH and OGP approaches are able to overcome the limitations previously mentioned, see Figures \ref{fig:bridge_koh_no_temp} and \ref{fig:bridge_ogp_no_temp}. In particular, the bias-corrected response in both cases fits exactly the MSE of the observations at each load position. Due to the flexible bias term, the corrected prediction is not limited to representing a null displacement at the end of the experiment, allowing for a better representation of the observations. Remarkably, the fitted parameter $E$ obtained with KOH's approach converges with almost no uncertainty towards the real value of 40 GPa, while for OGP it tends to 57.9 GPa, a similar value to the one for the approach without bias. This disparity can be explained by observing the behaviour of the bias predictions for different values of $E$, as represented in Figure \ref{fig:bridge_bias_no_temp_comparison}. For the case of KOH's approach, the optimization objective is the value of $E$ with the "best-fitting bias", which translates in this case to a minimization of the marginal log-likelihood of the bias GP for a given $E$. Following Equation \ref{eq:gp_loglike}, large residuals and large covariance values are penalized. Observing Figure \ref{fig:bridge_bias_no_temp_comparison}, it can be appreciated that $E=40$ GPa generates a very smooth set of residuals due to the linear nature of the temperature increase over time, which greatly reduces the values of $C_b$. The reduction in the residuals for values with larger $E$ does not compensate for the increase in variance, which results in KOH eventually converging to a value close to $E=40$ GPa. It must be remarked that this is true for the chosen combination of kernel and observations due to the regularity and smoothness of the bias introduced by the temperature difference but is not generally true for any system. As mentioned before, it is not possible to control a priori the optimality metric of KOH's approach for an arbitrary choice of priors. Nevertheless, bias from physical sources is expected to be relatively smooth, which can favour such an approach.

In comparison, the imposition of the orthogonality condition onto the bias GP kernel leads the distribution of $E$ to converge towards the MSE. It must be noted that the correction generated by the bias from OGP only fits the observations in the case of an optimal value of $E$. The orthogonality condition restricts, through a modification of the covariance matrix, the function space that can represent the bias correction. In this case, as the derivative of the computational model with respect to the Young's modulus does not change much in the studied range of $E$, the orthogonal kernel remains almost constant for every evaluation. In particular, the functions generated by the bias GP will resemble the case where $E$ is optimal, the MSE of the observations. Additionally, the orthogonal kernel translates into an impossibility of the bias GP to be trained to fit residuals that would be improved by a modification of the latent parameter.For an arbitrary $E$, the bias will be able to correctly compensate for the discrepancy at the end of the experiment (which can only be explained through the bias) but not the values when the load is at the center of the span (which can be largely altered by a modification in the value of $E$). These diminished effects of the bias on some parts of the domain imply larger covariance values due the inability to cover the observed residuals with members of the imposed function space. The increased loss of smoothness due to high variance for values of $E$ far from 57.8 GPa can be observed in Figure \ref{fig:bridge_bias_no_temp_comparison}. Larger variations in the derivative of the computational model with respect to $E$ would lead to more dissimilar realizations of the bias response mean. Nevertheless, the convergence to the optimum $E$ is achieved thanks to diminishing covariances obtained by the fulfilment of the orthogonality condition in values closer to the MSE. The mean of the bias response corresponds with the residuals for that value in the whole domain of application.

Nevertheless, the predictions obtained through the approaches with bias still present some limitations. First, the system cannot reproduce the observations and only performs a regression on the different experimental sets. Second, if the bias had a more complex non-linear behaviour with respect to increments in temperature difference between series, the bias would not be able to reflect it. Finally, the obtained bias does not provide insight into the quality of the model with respect to the temperature differences at the end of the experiment, as it only corrects the displacement over time to achieve the MSE of the training observations, not considering the end difference temperature for the associated experiment. These shortcomings can be mitigated by extending the input of the domain for the bias GP.

\subsubsection{Models with temperature consideration}

A possible alternative to cover the temperature variations in the system is to include the noise standard deviation $\sigma_\epsilon$ in the latent parameters instead of prescribing it, as it was presented in Section \ref{sub:cantilever_beam} for the application without bias term. However, this does not take advantage of the temperature information nor increases the model's predictive capacity. 

Alternatively, we make the bias GP depend on the temperature difference as well $b(x)\to b(x,\Delta T)$. Its main interest resides in the inclusion of information that may be relevant to the process without modifying the computational model itself. This extension of the domain of interest through the bias allows weighting in the influence of the temperature difference in the inference process while providing insight into the potential model deficiencies. Therefore, the bias GP will be trained on pairs of the form $(x,\Delta T)$, where $x$ is the current load position and $\Delta T$ is the associated end temperature difference for a given observation.  This bias can consequently be represented by a response surface with a 2D input and a 1D output. Each series of observations is part of the same experiment with an associated $\Delta T$ measured at the end of it. The intermediate temperatures are not available, therefore all the observations will share the final $\Delta T$ as their coordinate in the extended domain. The training of the bias term and the inference procedures are implemented following the algorithms from Section \ref{sec:model_bias_approaches}. In this case, an additional scaling of the domain and the residuals is applied to improve the fitting of the bias GP. The GP inputs are scaled such that the original domain $(x,\Delta T):[0,104]\times[0.01,0.1]$ is transformed to $[0,1]\times[0,1]$. The residuals and derivatives are multiplied by a factor of 100 during the training to increase numerical stability. The originally small residuals produced by the displacements lead to values of the variance close to machine precision after scaling the time domain, which leads to numerical instabilities of the GP training. To compensate this effect, the scaling factor in the residuals is chosen in the same order of magnitude as the original time length. The same previous kernels are applied for KOH and OGP, extending them to accept 2D inputs.

\begin{figure}[ht]
	\centerline{\includegraphics[width=\textwidth]{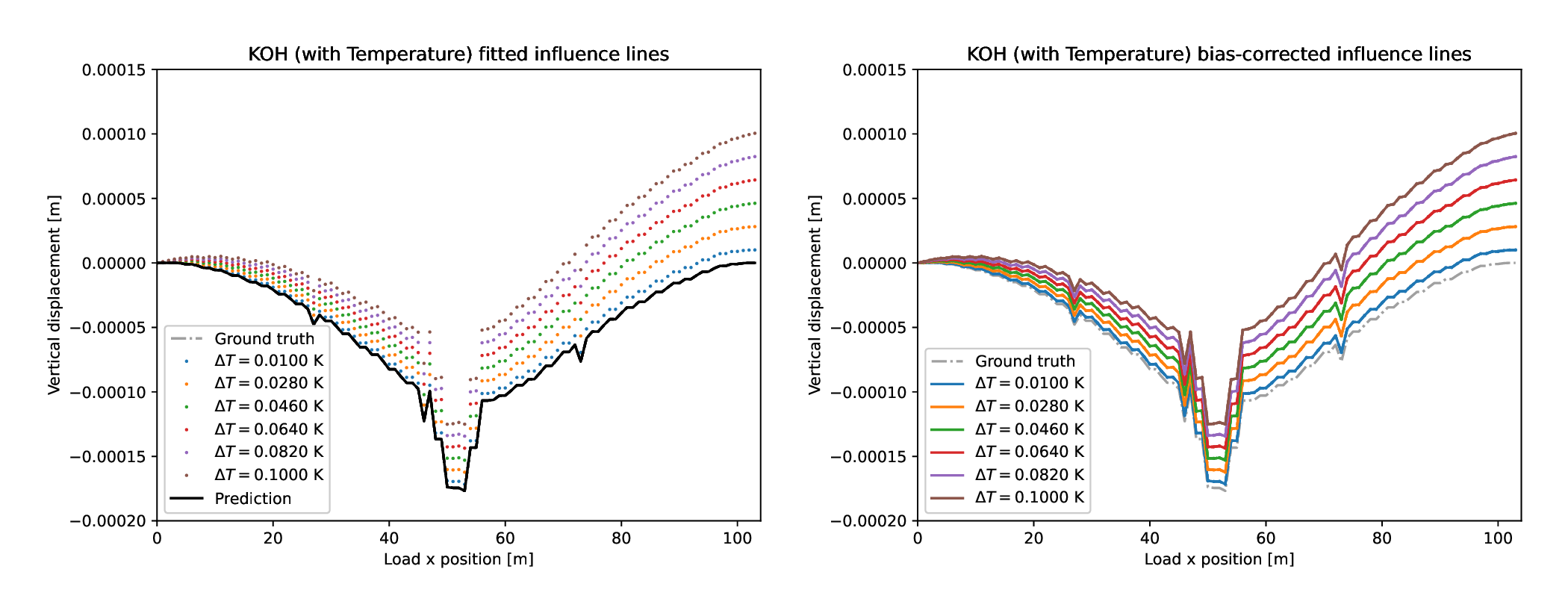}}
	\caption{Predicted response of the fitted model using KOH's approach including the temperature information. Left, fitted response of the computational model. The prediction coincides almost exactly with the ground truth, and its variance is not visible. Right, bias-corrected response for each temperature series.}
	\label{fig:bridge_koh}
\end{figure}

\begin{figure}[ht]
	\centerline{\includegraphics[width=\textwidth]{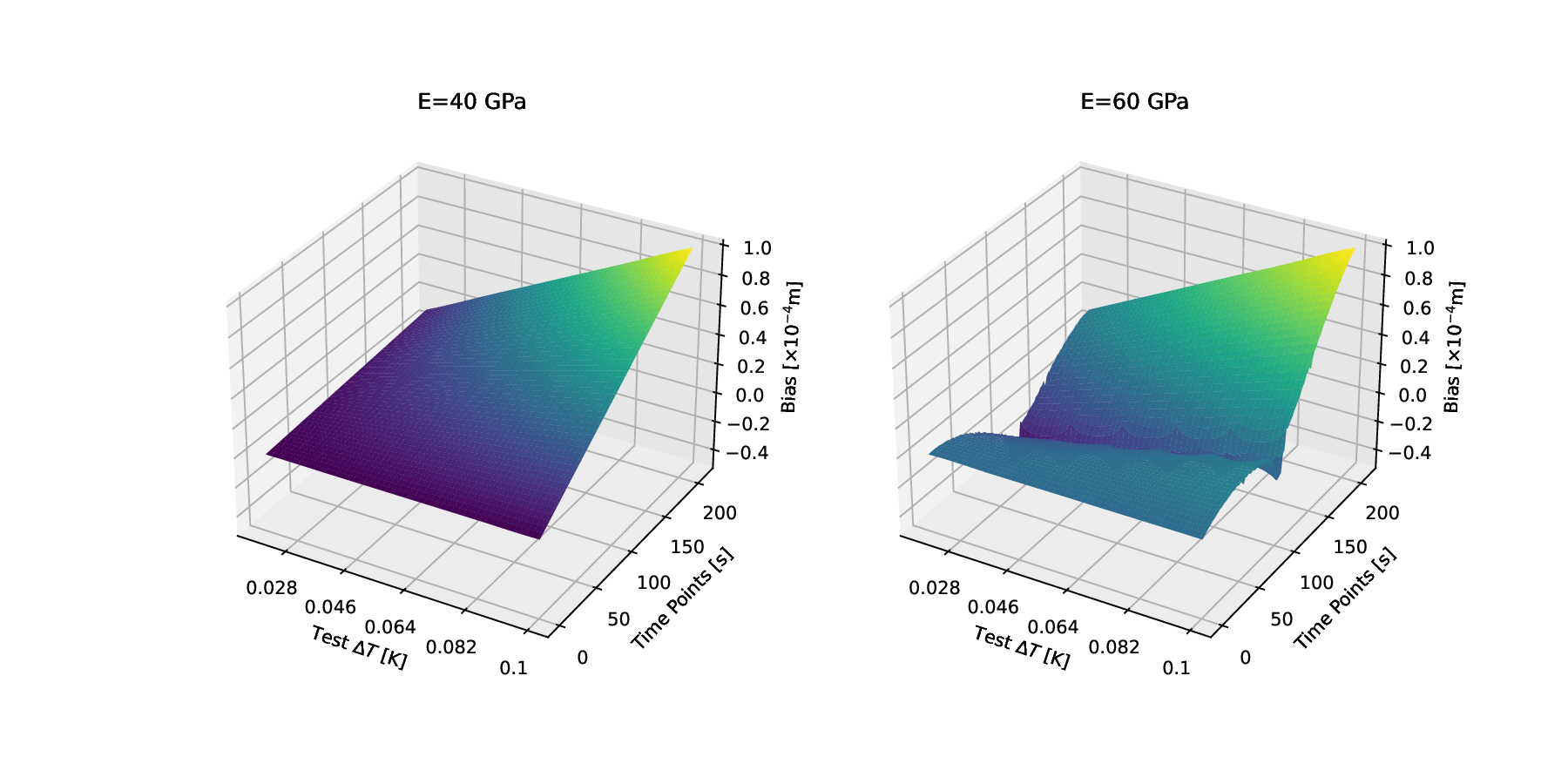}}
	\caption{Comparison of bias response surfaces for KOH with $E$=40 GPa and $E$=60 GPa.}
	\label{fig:bridge_koh_bias_comparison}
\end{figure}

\begin{figure}[ht]
	\centerline{\includegraphics[width=\textwidth]{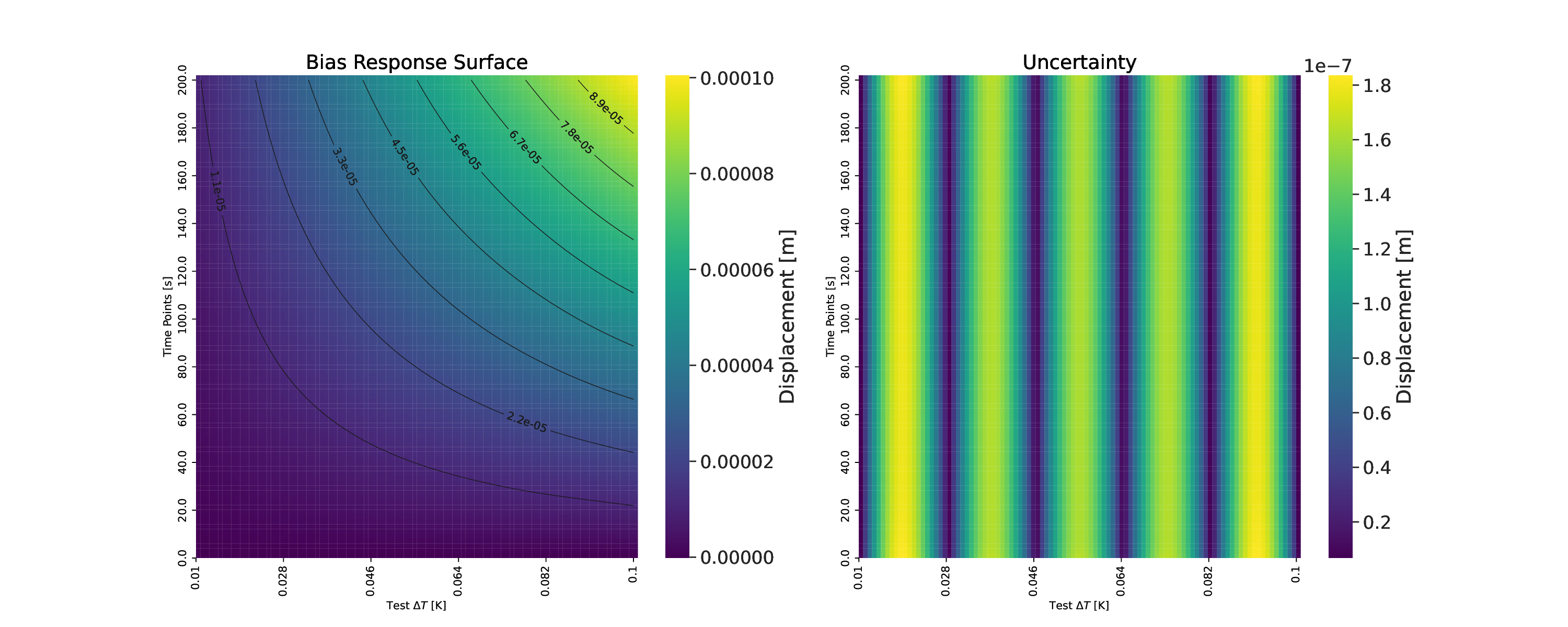}}
	\caption{Bias plots for KOH approach for the fitted bias for the MLE $E=40$ GPa. Bias GP response surface (left) and its associated variance (right).}
	\label{fig:bridge_koh_bias_plots}
\end{figure}

% KOH approach
The results of including temperature information in KOH's approach can be observed in Figure \ref{fig:bridge_koh}. Analogously to the case without temperature, the inferred $E$ tends to 40 GPa due to the linearity in the bias with respect to the temperature increments and in time. A comparison of the bias response surface for $E$=40 GPa and $E$=60 GPa is presented in Figure \ref{fig:bridge_koh_bias_comparison}. As in the previous evaluation, the reduction of the residuals provided by larger values of $E$ does not compensate for the increase in complexity, which is translated to larger values for the determinant of the covariance matrix and, consequently, smaller likelihoods. However, if the predictions are corrected by the learnt bias term trained on the MLE of $E$, it is possible in this case to regenerate the observations for each temperature gradient. This corrected model can be used for generating influence lines for other temperature gradients in the training range. Nevertheless, the inferred bias cannot be trusted for other applications, as it is defined only on the predicted influence lines, but it can be useful for identifying trends in the model discrepancy. An analysis of the bias contour lines and their variance such as the one presented in Figure \ref{fig:bridge_koh_bias_plots} points out increasing uncertainties with larger temperature differences and at the end of each experiment. Additionally, the largest variances occur between samples with different temperature increments. These observations coincide with the expected behaviour of the model. If a noise component were to be inferred additionally, the variance would be significantly reduced in comparison with the case without temperature awareness, as the observations would be considered independently and not as realizations of the same distribution.


\begin{figure}[ht]
	\centerline{\includegraphics[width=\textwidth]{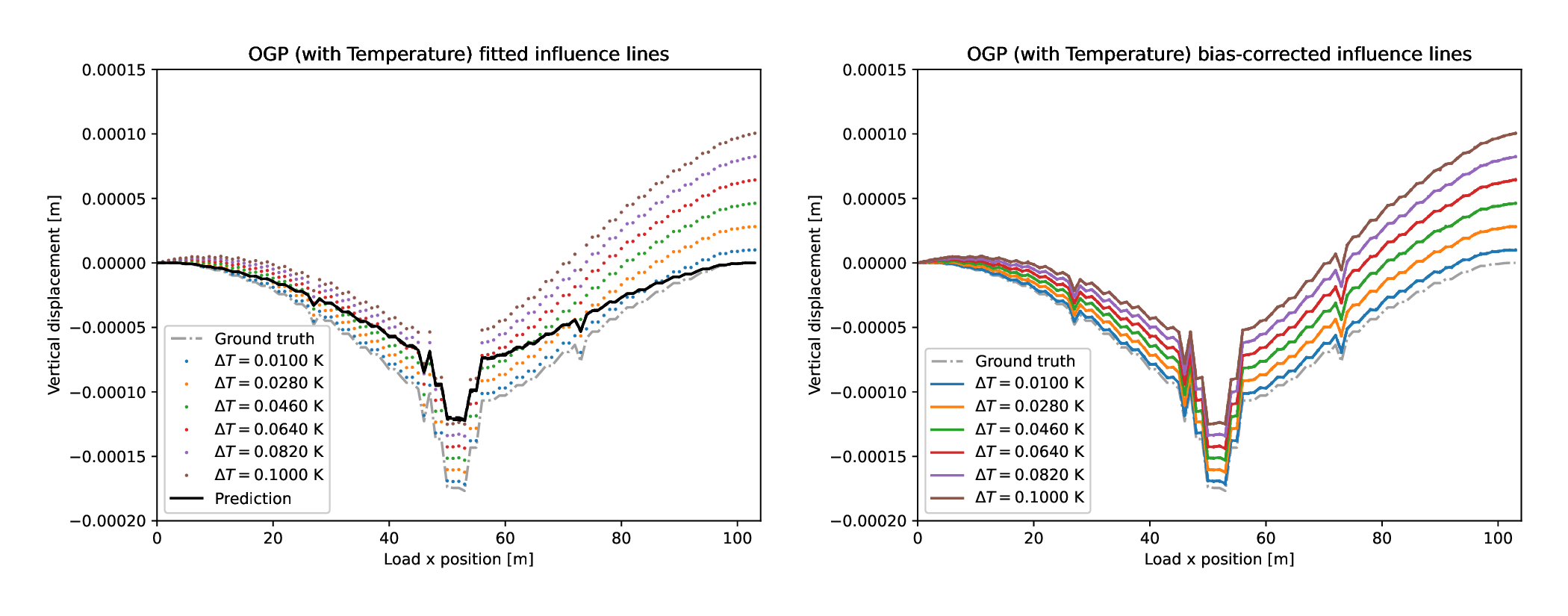}}
	\caption{Predicted response of the fitted model using OGPs including the temperature information. Left, fitted response of the computational model. The prediction's variance is not visible. Right, bias-corrected response for each temperature series.}
	\label{fig:bridge_ogp}
\end{figure}

\begin{figure}[ht]
	\centerline{\includegraphics[width=\textwidth]{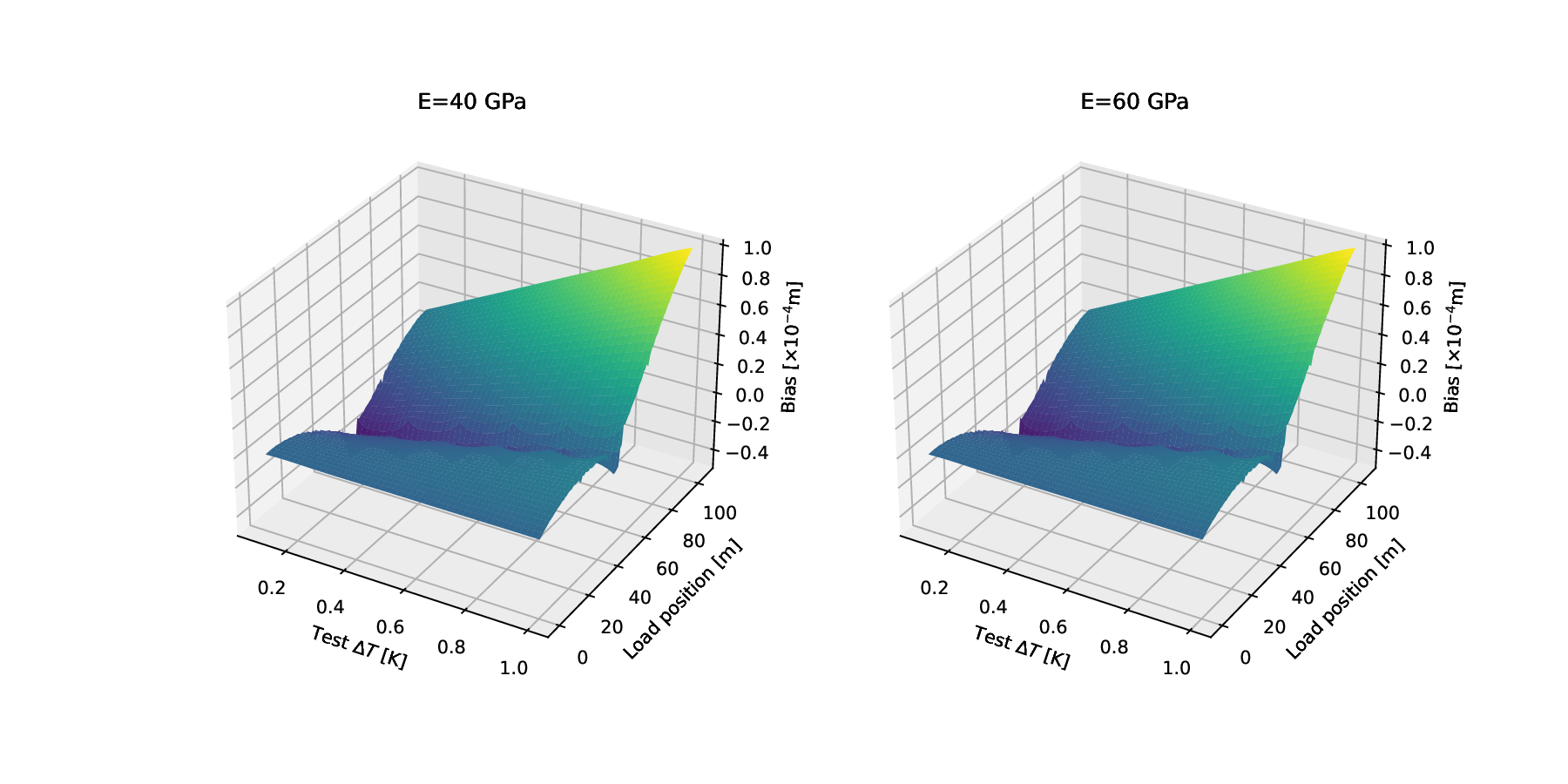}}
	\caption{Comparison of bias response surfaces for OGP with $E$=40 GPa and $E$=60 GPa.}
	\label{fig:bridge_ogp_bias_comparison}
\end{figure}

\begin{figure}[ht]
	\centerline{\includegraphics[width=0.9\textwidth]{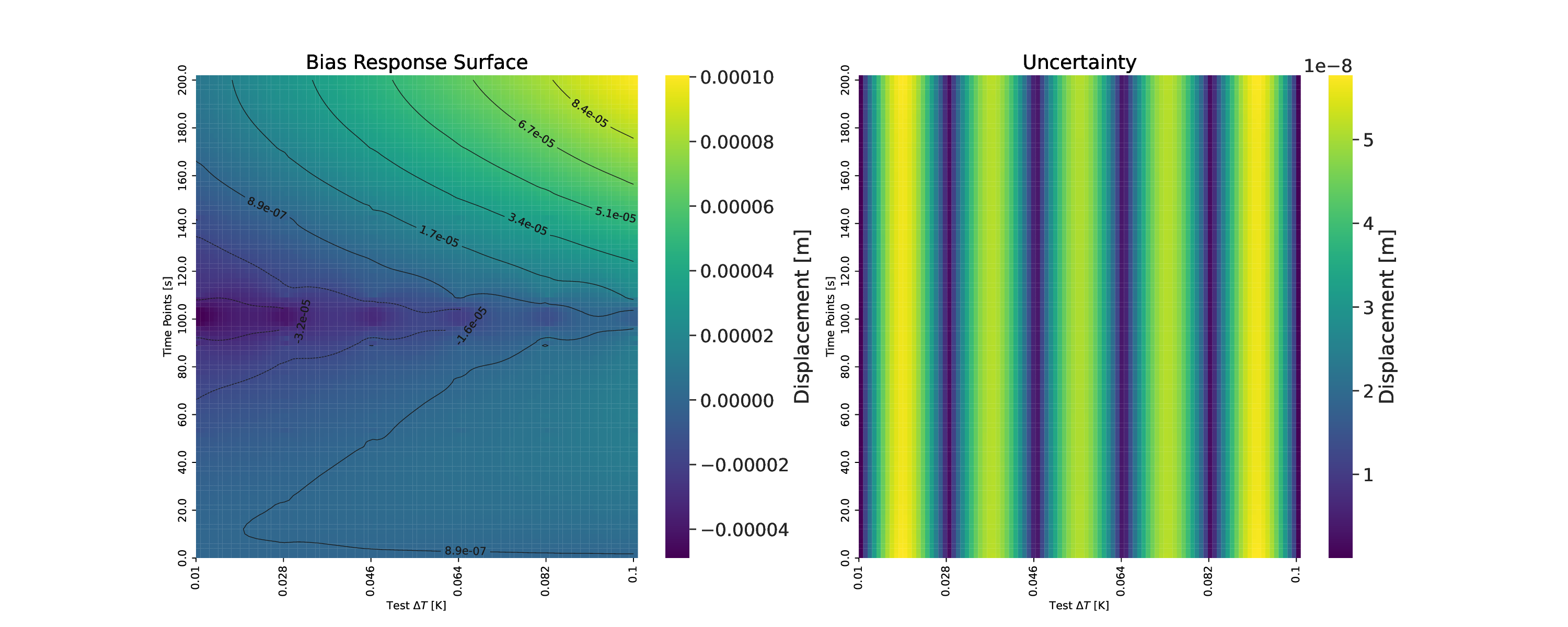}}
	\caption{Bias plots for OGP approach. Bias GP response surface (left) and its associated variance (right).}
	\label{fig:bridge_ogp_bias_plots}
\end{figure}

% OGP approach
Finally, the influence lines and corrected predictions for OGP's approach are represented in Figure\ref{fig:bridge_ogp}. The orthogonal prior imposed in OGP's approach favours convergence to the sampled $E$ that minimizes the prescribed loss between computational and generative models, the $L^2$ loss in this case. As the computational model is not aware of the temperature gradient, the observations are sufficiently detailed and the anchor points of the computational model coincide with the observation points, minimizing such a loss is almost equivalent to minimizing the MSE, analogously to the case where temperature is not included in the bias. Consequently, the optimal value of $E$ obtained by imposing an OGP on the bias coincides with the best-fitting computational model to the observation but disregards the additional information introduced by the temperature gradient. In this case, this means a tendency towards 57.8 GPa. The comparison of the bias response surfaces represented in Figure \ref{fig:bridge_ogp_bias_comparison} presents an analogous behaviour as the case where the temperature is not included, where the fitness of the GP is driven by its variance due to the small changes in the derivative of the computational model with respect to $E$ in the chosen domain. It is remarkable that, as shown in Figure \ref{fig:bridge_ogp_bias_plots}, the bias response surface presents relatively low uncertainty, despite not being able to fit the observations. This surface informs about the distribution of the discrepancy between model prediction at the MSE and the measurements, which could be useful to identify regions of low certainty in the input parameter space, indicating deficiencies in the model. Nevertheless, as the inference converges to the optimum value of $E$, the bias-corrected predictions are able to reflect the observations, including the temperature effect.


In comparison, KOH and OGP outperform the classical approach with and without considering the temperature increment in the bias. If the model does not include information about the temperature, the bias correction allows to regenerate the MSE of the observations and provides insight into the evolution of the discrepancy depending on the load's position. However, the inclusion of the temperature differences in the bias informs of deficiencies in the model with respect to them, as well as the distribution of the uncertainty. Remarkably, this is achieved without needing to modify the computational model, only making use of already available information. The insight provided by the inclusion of temperature can be valuable for the design of future measurement campaigns or for the identification of potential improvements in the computational model, which are of great relevance in the context of digital twins. Additionally, if the computation times presented in Table \ref{tab:times_bridge} are considered, the cost of implementing KOH's approach to extend the model to include the temperature difference is negligible in comparison with the approach without temperature or even without bias. Most of the computational time is spent in evaluating the model to generate the influence lines for each sample while training the GP on $(x,\Delta T)$ as needed for KOH's approach is faster in comparison. OGP's additional time comes from the necessity to solve the system several times per evaluation to calculate the gradient. A more efficient differentiation approach would reduce the difference in time between OGP and KOH. In conclusion, KOH's approach should be applied when possible to extend the system through the bias due to its reduced computational cost, simple integration and valuable insights. It is possible that for more complex cases, where more than one extra coordinate is introduced, this approach may incur further identifiability issues.

Another relevant consideration is the capacity of the predicted response to fit the observations. The Mahalanobis distance can be used to compare the posterior distribution of the bias-corrected responses for each approach with the observations. As the posterior distributions are relatively narrow, the Mahalanobis distances is almost equivalent to the Euclidean one, while the latter is numerically easier to compute due to the absence of a badly conditioned inverse matrix. The Euclidean distance for each approach with respect to the observations is presented in Table \ref{tab:times_bridge}. The approaches with bias and no temperature information show slight improvement over the approach without bias due to their capacity to provide predictions out of the limits of the computational model, such as the zero displacement at the end of the experiments. However, the largest improvement comes from the inclusion of the temperature differences in the bias, as these approaches are able to fit every experiment series. In the absence of prior information, these distances can be used to assess the impact of introducing the temperature differences in the system and to guide further refinements of the digital twin simulation model. As expected, there is a clear benefit in this case for introducing the available temperature information.

\begin{table}[ht]
	\centering
	\begin{tabular}{ccc}
		\hline
		Method  & \multicolumn{1}{c}{Time} & \multicolumn{1}{c}{Euclidean distance} \\ \hline
		No bias & 5 h 15 min        & 4.77e-4                  \\
		KOH without temperature & 5 h 21 min & 4.49e-4\\
		KOH with temperature & 5 h 30 min      & 9.41e-9    \\
		OGP without temperature & 15 h 15 min & 4.46e-4\\
		OGP with temperature  & 15 h 31 min & 1.57e-6 \\ \hline
	\end{tabular}
	\caption{Evaluation times and corresponding Euclidean distances for the third benchmark (bridge case).}
	\label{tab:times_bridge}
\end{table}

\section{Conclusion}
\label{sec:conclusion}
In this paper, we have proposed solutions to some of the main challenges for the reliable implementation of simulation-based digital twins of bridges. A Bayesian inference framework is often considered the best option for the calibration of the model parameters. However, we observed that classical approaches led to overconfident models whose response does not replicate the observations and whose behaviour does not correspond to the physical system. This renders them unsuitable for their use in digital twins of critical infrastructures such as bridges, where fidelity, robustness and accuracy are key to guarantee their safety. The root of this inadequacy stems from the inability of computational models to recreate the infinitely complex reality to its fullest extent. We proposed the explicit introduction of a model bias term in the formulation and calibration of the digital twin to mitigate such an effect. Two different approaches were compared with the classical inference methodology, one with a modularized version of Kennedy and O'Hagan's framework and one based on OGPs. Through three relevant use cases, we proved that biased approaches did not only compensate for the model bias but also provide new insight into model deficiencies. Additionally, we introduced two novel extensions to the model bias identification schemes to correctly quantify noise in the observations from unknown physical sources and to make use of complementary information. The bias identified through these extensions can be used as a bases for guiding potential model improvements.

First, when applied to a simple 1D case where model bias is present, the classical approach converged to a suboptimal parameter value with high certainty, despite not being able to fit the observations. Contrarily, both biased approaches were able to correct successfully the predictions to fit the observations, properly address the uncertainty and, for the case of OGP, to provide the optimal value for the latent parameter. In the context of digital twins, the introduction of the orthogonality condition allows for the use of fewer data point locations, i.e. sensors, while keeping control of the optimality of the solution. Additionally, the extension of the solution to the domain of the anchor points instead of the sensors generates reliable solutions at points of interest where no data is available. This is generally done at the expense of longer computational times and model evaluations at those points, which can be a challenge for complex and expensive simulation models. 

Next, the three methodologies were applied to a stochastic cantilever beam with model bias and additional noise from a non-prescribed physical source. Although the estimated Young's modulus differed from the real value in any case, only the biased approaches were able to estimate the noise and correct the predictions, but could not properly identify the noise due to the presence of model bias. Both KOH and OGP enable the introduction of homoscedastic and heteroscedastic noise kernels without modifications to the system or the inference procedure, which provides great flexibility. The introduction of such kernels has proven key to identifying noise under conditions of bias.

Finally, a test demonstrator example of the Nibelungenbrücke was implemented, where a discrepancy from a temperature difference was introduced to evaluate the ability of the approaches to include new information into the system. The extension of the model through the bias allowed us to gain relevant insight into the deficiencies in the model just by leveraging already available information. The computational cost of implementing KOH compared to not including the bias term is negligible for such a complex model, which makes this approach very promising for its implementation in digital twins of bridges.

However, one of the main challenges of the presented methodologies is the inability to assure the physicality of the inferred parameters due to the addition of the corrective bias term at the output of the predictions. The model itself is therefore not directly modified, and the corrections cannot be utilized for the calculation of different quantities than the ones used for the calibration. Nevertheless, the generated bias and the corrected models allowed us to better understand the deficiencies present in the current model in a quantitative manner. Additionally, the evaluation of the presented frameworks under real measurements is required for validation of the exposed approaches in structures as-built. The complexity of sensor observations and the variation in environmental and operational conditions cannot be replicated with simulated data, therefore the evaluation of a definitive assessment of their performance is only possible under such conditions. Furthermore, the integration of these methodologies in a unified digital twin of a bridge presents additional challenges related to accessibility, ownership, connectivity and cybersecurity that should not be understated.

In conclusion, a suitably selected biased approach outperformed classical implementations in every case, proving their effectiveness in the reduction of the effects associated with the model bias. OGPs provide optimal results at the expense of increased computational times when the model already considers the domain of interest, while KOH's approach allows for the extension of such domain with additional information. The choice of one or the other will depend on the complexity of the model and its intended use.

% Here's the list of references:
%
\label{section:references}
\bibliography{my_bibliography}
%

%\backmatter
%\bmsection*{}
\section*{Author contributions}

\textbf{Daniel Andrés Arcones}: methodology (lead), formal analysis (lead), software(lead), visualization (lead), original draft preparation (lead). \textbf{Martin Weiser}: conceptualization (equal), supervision (equal), review and editing (equal). \textbf{Faidon-Stelios Koutsourelakis}:  conceptualization (equal), supervision (equal), review and editing (equal). \textbf{Jörg F. Unger}: conceptualization (equal), supervision (equal), review and editing (equal).

%\bmsection*{}
\section*{Acknowledgments}
BAM is a senior scientific and technical federal institute with responsibility to the German Federal Ministry for Economic Affairs and Climate Action.

%\bmsection*{}
\section*{Financial disclosure}

This work was funded through the project ``C07 - Data driven model adaptation for identifying stochastic digital twins of bridges'' from the Priority Program (SPP) 2388/1 ``Hundred plus'' of the German Research Foundation (Deutsche Forschungsgemeinschaft, DFG).

%\bmsection*{}
\section*{Conflict of interest}

The authors declare no potential conflict of interests.

%\bmsection*{}
\section*{Supporting information}

Additional supporting information may be found in the online version of the article at the publisher’s website.

%\nocite{*}% Show all bib entries - both cited and uncited; comment this line to view only cited bib entries;

\end{document}